\documentclass[11pt]{article}
\usepackage{epstopdf}                                                             
\usepackage[a4paper,hmarginratio=1:1,vmarginratio=2:3,totalwidth=15.4cm,totalheight=22.56cm]{geometry}
\usepackage{bm,epstopdf,epsfig,amsmath,amssymb,amsfonts,colordvi,latexsym,wrapfig,comment,cancel,verbatim,slashed}
\usepackage{epsfig,bbm}
\usepackage{graphicx,graphics}
\usepackage[font=md,captionskip=8pt]{subfig}
\usepackage[usenames,dvipsnames]{color}
\usepackage[noadjust]{cite}
\usepackage{xcolor} 
\usepackage[utf8]{inputenc}
\usepackage{setspace}

\usepackage[utf8]{inputenc}

\newcommand{\q}{\alpha}

\newcommand{\sg}{\sqrt{g}}    
\newcommand{\sgh}{\sqrt{\hat g}}
\newcommand{\w}{\omega}

\allowdisplaybreaks

\newcommand{\si}{{\sigma}}
\newcommand{\cA}{{\cal A}}

\newcommand{\cL}{{\cal L}}
\newcommand{\cM}{{\cal M}}

\newcommand{\cO}{{\cal O}}   
\newcommand{\cP}{{\cal P}}

\newcommand{\cZ}{{\cal Z}}

\newcommand{\ra}{\rightarrow}

\newcommand{\be}{\begin{equation}}
\newcommand{\ee}{\end{equation}}
\newcommand{\bea}{\begin{eqnarray}}
\newcommand{\eea}{\end{eqnarray}}

\newcommand{\baa}{\begin{array}}
\newcommand{\eaa}{\end{array}}

\long\def\symbolfootnote[#1]#2{\begingroup
\def\thefootnote{\fnsymbol{footnote}}\footnote[#1]{#2}\endgroup}

\begin{document} 
\begin{flushright}
  % \today\\
\end{flushright}
\bigskip\medskip
\thispagestyle{empty}
\vspace{2cm}

\begin{center}
\vspace{0.5cm}

 {\Large \bf   Standard Model in Weyl conformal geometry} 

 \vspace{1.5cm}
 
 {\bf D. M. Ghilencea}$^a$
 \symbolfootnote[1]{E-mail: dumitru.ghilencea@cern.ch}
 
\bigskip

$^a$ {\small Department of Theoretical Physics, National Institute of Physics
 \smallskip 

 and  Nuclear Engineering (IFIN), Bucharest, 077125 Romania}
\end{center}

\medskip

\begin{abstract}
  \noindent
  \begin{spacing}{1.}
    \noindent
We study the Standard Model (SM)  in  Weyl conformal geometry. This  embedding
is {\it natural} and  truly minimal {\it with  no new fields} required beyond the SM spectrum
and  Weyl  geometry.    The action inherits a  {\it gauged} scale symmetry $D(1)$
(known as Weyl gauge symmetry) from the underlying geometry.
The associated Weyl quadratic gravity undergoes  spontaneous breaking of
$D(1)$  by a geometric Stueckelberg mechanism in which the Weyl gauge field ($\w_\mu$)
acquires mass by ``absorbing'' the spin-zero mode ($\phi$) of the $\tilde R^2$ term in the action.
This mode also generates the Planck scale and the cosmological constant.
The  Einstein-Proca action of $\omega_\mu$ emerges in the broken phase.  In the presence of the SM,
this mechanism receives  corrections (from the Higgs) and it  can induce
electroweak (EW) symmetry breaking.
The EW scale is proportional  to the vev of the Stueckelberg field ($\phi$).
From the SM spectrum, only the Higgs field ($\sigma$) has direct couplings  to the Weyl gauge field,
and its mass  may be protected  at  quantum level by the $D(1)$ symmetry.
The  SM fermions can acquire couplings to $\w_\mu$ only in the special case of a
non-vanishing        kinetic mixing of the gauge fields of  $D(1)\times U(1)_Y$.
If this mixing is indeed present, part of $Z$ boson mass 
is not due to the Higgs mechanism, but to its  mixing with massive $\w_\mu$.
Precision measurements of $Z$ mass then set lower bounds on the mass of $\w_\mu$
which can be light (few TeV).
In the early Universe the Higgs can have a {\it geometric} origin  by Weyl vector fusion
and the Stueckelberg-Higgs potential  can drive inflation.
The  dependence of the tensor-to-scalar ratio $r$ on the
spectral index $n_s$  is  similar to that in Starobinsky inflation but
shifted to lower $r$ by the Higgs non-minimal coupling to Weyl geometry.
\end{spacing}
\end{abstract}

\newpage

\section{Motivation}

The Standard Model (SM)  with the  Higgs mass parameter set to zero has a scale symmetry.
This may indicate that this symmetry plays a role in model building for physics beyond the SM \cite{Bardeen}.
Scale symmetry is natural in physics at higher scales or in the early Universe
when all states are essentially  massless. In such scenario, the mass terms and scales
of the theory e.g. the Planck and electroweak (EW)
scales must be generated by the  vacuum expectations
values (vev) of some scalar fields.
In this work we consider the SM with a {\it gauged} scale symmetry (also called Weyl
gauge symmetry) \cite{Weyl1,Weyl2,Weyl3} which we prefer to the more popular {\it global} scale
symmetry, since the latter is broken by black-hole physics \cite{Kallosh}.
A natural  framework for this symmetry is Weyl
geometry \cite{Weyl1,Weyl2,Weyl3} where this symmetry is {\it built in}.
We thus consider the SM embedded in the Weyl conformal geometry and study the implications.

The Weyl geometry is defined by classes of equivalence
$(g_{\alpha\beta}, \w_\mu$) of the metric ($g_{\alpha\beta}$)
and the Weyl gauge  field ($\w_\mu$),  related by the  Weyl gauge
transformation, see $(a)$ below. If matter is present, $(a)$ must be extended by
transformation  $(b)$ of the scalars ($\phi$) and fermions ($\psi$)
\bea\label{WGS}
 (a) &\quad&
 \hat g_{\mu\nu}=\Sigma^d %\Omega^2
 \,g_{\mu\nu},\qquad
\hat\w_\mu=\w_\mu -\frac{1}{\q}\, \partial_\mu\ln\Sigma, %\Omega^2
\qquad
\sqrt{\hat g}=\Sigma^{2 d} \sqrt{g},
\nonumber\\[5pt]
(b) &\quad & \hat \phi = \Sigma^{-d/2} \phi, \qquad \hat\psi=\Sigma^{-3d/4}\,\psi,
\qquad\qquad\quad (d=1).
\eea
Here $d$ is the Weyl charge of $g_{\mu\nu}$,
$\alpha$ is the Weyl gauge coupling\footnote{Our convention
  is  $g_{\mu\nu}=(+,-,-,-)$ %$g=\vert\det g_{\mu\nu}\vert$
  while the curvature tensors are defined as in  \cite{book}.},
$g=\vert\det g_{\mu\nu}\vert$ and $\Sigma>0$.
This is a non-compact gauged dilatation symmetry, denoted   $D(1)$.
Since it is Abelian, the  normalization of the charge $d$
is not fixed\footnote{For example  $d=1$  is a convention used in e.g.
  \cite{Smolin} while $d\!=\!2$ was considered in \cite{Kugo}.}.
In this paper we  take $d=1$. 
The case of arbitrary $d$ is recovered from our results  by simply  replacing  $\q\ra d\, \q$.
A discussion on  symmetry (\ref{WGS}) and a brief introduction to  Weyl geometry are found in
Appendix~\ref{AA}.

To study  the SM in Weyl geometry, all  one needs
to know for the purpose of this work  is the expression of the
connection ($\tilde\Gamma$)  of this geometry, which
differs from the Levi-Civita connection ($\Gamma$)  of (pseudo-)Riemannian case
used in Einstein gravity. The Weyl connection is a solution to
$\tilde\nabla_\lambda  g_{\mu\nu}=-  \q\, \w_\lambda g_{\mu\nu}$
where $\tilde\nabla_\mu$ is defined by  $\tilde\Gamma_{\mu\nu}^\lambda$.
This solution  is (see Appendix)
% \smallskip
\bea\label{tGamma}
\tilde \Gamma_{\mu\nu}^\lambda=
\Gamma_{\mu\nu}^\lambda+(1/2)\,\q \,\Big[\delta_\mu^\lambda\,\, \w_\nu +\delta_\nu^\lambda\,\, \w_\mu
- g_{\mu\nu} \,\w^\lambda\Big].
\eea
$\tilde\Gamma$  is invariant under (\ref{WGS}),  as it should be, since the  parallel
transport of a vector must be gauge independent. Taking the trace in (\ref{tGamma}), with a notation
$\tilde\Gamma_\mu=\tilde\Gamma_{\mu\nu}^\nu$ and $\Gamma_\mu=\Gamma_{\mu\nu}^\nu$, then
\be
\w_\mu \propto \tilde\Gamma_\mu-\Gamma_\mu.
\ee
The Weyl  field is thus a measure of the (trace of the) deviation from a Levi-Civita connection.

The general quadratic gravity action  defined by
Weyl geometry \cite{Weyl1,Weyl2,Weyl3}, invariant under (\ref{WGS}),
is written in terms of scalar and tensor curvatures of this geometry.
Using $\tilde\Gamma$ of (\ref{tGamma}) and standard formulae
one can express  these curvatures  in terms of their Riemannian counterparts
 and re-write the action  in a more  familiar Riemannian notation (as we shall do).
 In the limit  $\w_\mu\!=\!0$ i.e. if:  i) $\w_\mu$ is `pure gauge'
 % (non-dynamical)
 or if  ii)  $\w_\mu$  becomes  massive and decouples, 
 then  $\tilde\Gamma\!=\!\Gamma$ and then {\it  Weyl geometry becomes Riemannian!} This is
 an interesting transition, relevant  later.
 In i) invariance  under (\ref{WGS}) reduces to
 local scale invariance (no $\w_\mu$).

The role of Weyl gauge symmetry 
in model building beyond SM was  studied before \cite{Dirac,Kugo,
  Smolin,Cheng,Fulton,Wheeler,Moffat1,Nishino,Ohanian,Moffat2,Tann,
  ghilen,Guendelman,Quiros1,Quiros2,pp1,pp2,pp3,pp4,pp5}.
We go beyond these models which were limited to actions 
{\it linear} in the scalar curvature $\tilde R$ of Weyl geometry
and  which introduced {\it additional states} (scalar fields beyond the Higgs field)
to maintain  symmetry (\ref{WGS})  and to generate the mass scales (Planck, etc) of the theory.

Our approach here to model building
is truly {\it minimal}, in the sense that {\it  no new fields} are needed or
added to the SM spectrum  - we simply embed the SM in Weyl geometry! Note that the Weyl gauge field present here 
is  part of the underlying  geometry and of Weyl gravity\footnote{The
  literature often calls  Weyl gravity the square of the Weyl
  tensor in Riemannian geometry. We actually consider the  original Weyl quadratic gravity
  in Weyl geometry which has additional terms (Section~\ref{s2.1}).}.
The gravity part of the action
is fixed by  the Weyl geometry \cite{Weyl1,Weyl2,Weyl3},  is actually {\it quadratic}
and is automatically invariant under (\ref{WGS}) a)  (since $\tilde \Gamma$ is invariant).
This minimal approach  builds on our recent results  in \cite{Ghilen1}
(also \cite{Winflation,Winflation2,Ghilen2,Ghilencea:2022lcl})
that showed that the original Weyl quadratic gravity action {\it in the absence of matter}
is broken spontaneously to the  Einstein-Proca action. Thus,
{\it this breaking is geometric in nature}, no scalar field is added to this purpose
\cite{Ghilencea:2022lcl}.

 With this result,  embedding the SM in Weyl  geometry is very natural:
 one sets the Higgs mass parameter to zero and `upgrades' the SM covariant derivatives,
 to respect  symmetry (\ref{WGS}) inherited from  Weyl geometry. 
 Thus, both the Lagrangian and its underlying geometry ($\tilde\Gamma$) have the same
 Weyl gauge symmetry. This is a  {\it unique}  feature,  not present in  models with local
 scale symmetry  based on  Riemannian geometry (i.e. with no  $\w_\mu$). It
 adds mathematical consistency to the model and motivated this study.
 Hereafter we refer to this model as SMW.

There is additional motivation to study the SMW and the Weyl geometry:

\noindent
{\bf a)} Einstein gravity emerges naturally. After a Stueckelberg mechanism,
the Weyl gauge field $\w_\mu$  acquires a mass
$m_\w\sim \q\,M_p$ ($M_p$: Planck scale) by  ``eating'' the spin zero-mode ($\phi$)
of geometric origin  propagated by the  $(1/\xi^2)\tilde R^2$ term in the action of coupling $\xi$.
 The gauge fixing of symmetry (\ref{WGS}) is dynamical, as shown by the  equations of motion.
After $\w_\mu$
decouples, the Einstein action is naturally obtained as a broken phase of Weyl gravity.
$M_p$ and the cosmological constant ($\Lambda$) are
both generated by $\langle\phi\rangle$ and  are  related: $\Lambda/M_p^2= (3/2) \xi^2$.

\noindent
{\bf b)}  The theory has a symmetry  $D(1)\times U(1)_Y\times SU(2)_L\times SU(3)$.
Note that a gauge kinetic  mixing of $\w_\mu$  with the hypercharge field $B_\mu$  of $U(1)_Y$
is not forbidden by this symmetry.

\noindent
{\bf c)} The  Higgs field has couplings to $\w_\mu$, of type $\sigma^2w_\mu\w^\mu$. The
SM gauge bosons and fermions do not couple to $\w_\mu$   \cite{Kugo}.
Only if   a gauge kinetic mixing  exists, can fermions  couple to $\w_\mu$.

\noindent
{\bf d)} The SM  Higgs potential is recovered for  small Higgs field
values (relative to Planck scale).
The EW symmetry breaking is then induced  by gravitational effects, with the Higgs mass and
electroweak scale obtained for perturbative couplings of the Weyl quadratic gravity. 

\noindent
{\bf e)} 
If a gauge kinetic mixing is indeed present,
part of the $Z$ boson mass is not due to the Higgs mechanism, but 
to the geometric Stueckelberg mechanism (giving mass to $\w_\mu$).
Precision  data on $m_Z$ then constrains  the Weyl gauge coupling $\q$
and  the mass of $\w$.

\noindent
{\bf f)}  The Higgs potential at  large  field values drives inflation.
Interestingly, the origin of the Higgs field in the early Universe
is  geometrical, from the  Weyl boson fusion, see {\bf  c).}
The  prediction for the tensor-to-scalar ratio ($r$) (for  given
spectral index $n_s$) is  bounded from above by
that in the Starobinsky model with similar dependence $r(n_s)$,
due to the $\tilde R^2$ term.

\noindent
{\bf g)} The SMW can provide a successful  alternative to the $\Lambda$CDM, as discussed in \cite{CosmoSMW}.

These interesting properties of the SMW are studied  in Section~\ref{s2}.
 The relation to other scale-invariant models  follows (Section~\ref{s3}).
The Conclusions are in Section~\ref{s4}. The Appendix has 
 an introduction to Weyl conformal geometry and additional calculations for Section~\ref{s2}.

\section{SM in Weyl conformal geometry}\label{s2}

\subsection{Einstein action from spontaneous breaking of Weyl quadratic gravity}\label{s2.1}

Consider first the original  Weyl gravity action
\cite{Weyl1,Weyl2,Weyl3} and here we follow \cite{Ghilen1}. The action is
\medskip
\bea\label{inA}
\cL_0=\sg\, \,\Big[\, \frac{1}{4!}\,\frac{1}{\xi^2}\,\tilde R^2  - \frac14\, F_{\mu\nu}^{\,2} 
-\frac{1}{\eta^2}\,\tilde C_{\mu\nu\rho\sigma}^{\,2}\Big],
\eea
with couplings $\xi,\, \eta\leq 1$. Here
$F_{\mu\nu}=\tilde\nabla_\mu\w_\nu-\tilde\nabla_\mu\w_\nu$ is the field strength of $\w_\mu$,
with $\tilde\nabla_\mu\w_\nu=\partial_\mu\w_\nu-\tilde\Gamma_{\mu\nu}^\rho\w_\rho$.
Since  $\tilde\Gamma_{\mu\nu}^\alpha\!=\!\tilde\Gamma_{\nu\mu}^\alpha$ is symmetric,
 $F_{\mu\nu}=\partial_\mu\w_\nu-\partial_\nu\w_\mu$.
$\tilde C_{\mu\nu\rho\sigma}$ and $\tilde R$
are the  Weyl tensor and scalar curvature in  Weyl geometry, derived
from eq.(\ref{tGamma}). 
Their relations  to Riemannian Weyl-tensor $C_{\mu\nu\rho\sigma}$
and scalar curvature $R$ are shown in  eqs.(\ref{tR}), (\ref{tC}):
\bea%\label{def0}
\tilde C_{\mu\nu\rho\sigma}^2&=& C_{\mu\nu\rho\sigma}^2+ \frac32 \,\q^2\,F_{\mu\nu}^2,
\nonumber\\
\tilde R&=&R - 3\, \alpha\, \nabla_\mu\w^\mu - \frac32 \,\alpha^2\, \w_\mu \w^\mu.
\label{def}
\eea
% \medskip\noindent
The rhs of these equations is in a Riemannian notation, so
$\nabla_\mu\w^\lambda=\partial_\mu \w^\lambda+\Gamma^\lambda_{\mu\rho}\,\w^\rho$.

Each term in $\cL_0$ is invariant under $D(1)$ of  (\ref{WGS}). Indeed,
$\tilde R$ transforms  as ${\tilde R}\ra (1/\Sigma)\, \tilde R$ (see Appendix), so
$\sqrt{g}\, \tilde R^2$ is invariant. Also
$\sqrt{g}\ C_{\mu\nu\rho\sigma}^2$
and $F_{\mu\nu}^2\sqrt{g}$ are  invariant;
similar for $\sqrt{g} \,\tilde C_{\mu\nu\rho\sigma}^2$.
The term $\tilde C_{\mu\nu\rho\sigma}^2$  ensures that $\cL_0$ is
the  most general Weyl action,  but is largely spectator  under the transformations
below, so its impact can be analysed separately. Since it may be generated at  quantum level
 we included it (it brings a massive spin-2 ghost \cite{LAG}).

In $\cL_0$ we replace  $\tilde R^2\ra -2\phi^2 \tilde R-\phi^4$ with $\phi$ a scalar field.
Doing so gives a classically equivalent $\cL_0$, since by using the solution $\phi^2=-\tilde R$
of the equation of motion of $\phi$ in the modified $\cL_0$, one recovers action (\ref{inA}).
With eq.(\ref{def}), $\cL_0$ becomes in a Riemannian notation
\bea\label{alt}
\cL_0=\sqrt{g}\,\Big\{ \frac{-1}{12\,\xi^2}\,\phi^2\,\Big[ R- 3 \q \nabla_\mu\w^\mu -\frac{3}{2}
\q^2\,\w_\mu\w^\mu\Big]
-\frac{\phi^4}{4!\,\xi^2} - \frac14 \Big[ 1+ 6\frac{\q^2}{\eta^2}\Big] \,F_{\mu\nu}^2
-\frac{1}{\eta^2}\,C_{\mu\nu\rho\sigma}^2\Big\}
\eea

\smallskip\noindent
or, making the symmetry manifest
\smallskip
\bea\label{alt2}
\cL_0&=&\!\sqrt g\,
\Big\{\,\frac{-1}{2\xi^2} \,\Big[ \frac16\,\phi^2\,R \,+\,(\partial_\mu\phi)^2
-\frac{\q}{2}\nabla_\mu\,(\w^\mu\phi^2)
\Big]
-\frac{\phi^4}{4!\,\,\xi^2}\,
+
\,\frac{\q^2}{8\,\xi^2}\,\phi^2 \,\Big[\w_\mu-\frac{1}{\q}\partial_\mu \ln\phi^2\Big]^2
\!
 \nonumber\\[1pt]
 &&\qquad\qquad
-\frac{1}{4\,\gamma^2}\,F_{\mu\nu}^2\,-\,\frac{1}{\eta^2}\,C_{\mu\nu\rho\sigma}^2
\Big\}, \label{ST}
 \qquad \textrm{with}\qquad
1/\gamma^{2}\equiv 1+ 6 \,\q^2/\eta^2\geq 1.
\eea

\medskip\noindent
Every term of coefficient $\propto 1/\xi^2$ and the entire  $\cL_0$
are invariant under  (\ref{WGS});  we must then  ``fix the gauge'' of this symmetry.
This follows  from the equations of motion of $\phi$, $\omega_\mu$ (see later), while 
at the level of the Lagrangian this is
done by applying to $\cL_0$  a specific form of transformation (\ref{WGS}) that is scale-dependent
$\Sigma=\phi^2/\langle\phi^2\rangle$ which is fixing $\phi$ to its vev (assumed to exist,
generated e.g. at quantum level);
naively, one simply sets $\phi\ra \langle\phi\rangle$ in (\ref{ST}). In
terms of the transformed fields (with a ``hat''),  $\cL_0$ becomes
 \medskip
\be
\label{EP}
\cL_0=\sgh \,\Big[- \frac12 M_p^2\hat R +\frac34 M_p^2\,\q^2\,\gamma^2\hat\w_\mu \hat \w^\mu
- %\frac32\, \xi^2 M_p^4
\frac14\, \langle\phi^2\,\rangle M_p^2
-\frac{1}{4} \, \hat F_{\mu\nu}^2-\frac{1}{\eta^2}\,  C_{\mu\nu\rho\sigma}^2\Big],
\quad
M_p^2\equiv \frac{\langle\phi^2\rangle}{6\,\xi^2}.
\ee

\medskip\noindent
In (\ref{EP}) a total divergence in the action,
$\delta S=\q /(4\xi^2)\langle\phi^2\rangle \int d^4x \,\sqrt{\hat g}\,\nabla_\mu\hat\w^\mu$
was ignored - it may be replaced by a {\it local} condition  $\nabla_\mu\hat\w^\mu=0$.
This constraint will be obtained shortly from the  
current conservation of the symmetric phase, eqs.(\ref{alt}), (\ref{ST}).

In (\ref{EP}) we  identify $M_p$ with the Planck scale. Eq.(\ref{EP}) is the Einstein gauge (frame) and also
the unitary gauge of action (\ref{ST}).
By  a  Stueckelberg breaking mechanism  \cite{ST,P1,P2}, $\w_\mu$ has become a massive Proca field, 
after ``eating'' in (\ref{ST}) the derivative $\partial_\mu\ln\phi$ of the Stueckelberg field $\ln\phi$
\cite{Ghilen1}, which transforms with a shift under (\ref{WGS}).
It is important to note here that the
number of degrees of freedom (dof) is indeed conserved:
in addition to the graviton, 
the real, massless $\phi$ (dof=1) and
massless $\w_\mu$ (dof=2) were replaced by massive $\w_\mu$ (dof=3)
of mass $m_\w^2\!=\!(3/2) \q^2\gamma^2 M_p^2$ in eq.(\ref{EP}).
We shall see shortly that $\phi$ is indeed a dynamical field.
One may expect  $m_\w\!\sim\!  M_p$ but the {\it Weyl gravity coupling may naturally be
  $\q\! \ll\! 1$, so $m_\w\!\ll\! M_p$!}

The Einstein-Proca action in (\ref{EP}) is  a  broken phase of $\cL_0$ of (\ref{ST}).
After  $\w_\mu$ decouples from  (\ref{EP}), below $m_\w$  
the Einstein-Hilbert action  is obtained as
a `low-energy' effective  theory of Weyl gravity \cite{Ghilen1}.
Hence, Einstein gravity appears to be the  ``Einstein gauge''-fixed version of
the Weyl  action.  However, {\it the  breaking is more profound and is  not  the  result of
a mere  `gauge choice':}  it is  accompanied by
a  Stueckelberg mechanism and  by a  transition from  Weyl  to Riemannian geometry:
indeed, when massive $\w_\mu$ decouples then $\tilde\Gamma$ of\,(\ref{tGamma})\,is replaced\,by\,$\Gamma$.

In the other case, when  $\w_\mu$ is  light ($\q\ll 1$),
it  may be present in the action  at low energies, since the current non-metricity lower bound (set by the mass
$m_\w$ of $\omega_\mu$)
is actually very low, of few TeV only \cite{Latorre}!
It can also be a dark matter candidate e.g. \cite{Huang2,Tang}.

Note that the Stueckelberg term  in (\ref{ST})
\medskip
\bea
(\q^2/4)\,\phi^2\, \big[\,\w_\mu -(1/\q) \,\partial_\mu\ln\phi^2\big]^2=(\tilde D_\mu\phi)^2,\quad
\tilde D_\mu\phi\equiv \big[\partial_\mu -\q/2\,\,\w_\mu\big]\phi,
\eea

\medskip\noindent
is simply a Weyl-covariant kinetic term  of
the Stueckelberg field that became  the mass term of $\w_\mu$ in (\ref{EP}).
That is, a Weyl gauge-invariant  kinetic term of a (Weyl-charged) scalar
in Weyl geometry is a mass term for $\w_\mu$ 
in the (pseudo)Riemannian geometry underlying (\ref{EP}). This  gives an interesting
geometric interpretation to the origin of  mass, as a transition from Weyl to Riemannian
geometry,  without any scalar field present in the final spectrum.
The field  $\phi$ also generated the Planck mass and  was ``extracted''
from the $\tilde R^2$ term i.e. is of geometric origin (like $\w_\mu$), giving an elegant
breaking mechanism.

Further, from
(\ref{alt}) one writes the equation of motion of $\w_\mu$ and applying $\nabla_\rho$ to it,
one finds a conserved current (see  Appendix~\ref{apb} and \cite{Ghilen1})
\medskip
\be\label{jj}
J_\rho=-\frac{\q}{2\xi^2} \,\phi\, (\partial_\rho-\q/2\,\, \w_\rho) \phi, \qquad
\nabla_\rho J^\rho\!=\!0.
\ee

\medskip\noindent
This current conservation equation confirms  that $\phi$ is indeed a dynamical field,
which is relevant for the above Stueckelberg mechanism to take place.

This result also extends to the case of the gauged scale symmetry a
conserved  current $K_\rho=\phi\partial_\rho\phi$ in global  scale invariant theories,
with $\nabla^\rho K_\rho=0$.
For a Friedmann-Robertson-Walker (FRW) metric, 
$\nabla^\rho K_\rho=0$ had a solution $\phi(t)\ra$constant for large enough time ($t$),
so $\phi$ evolved to a vev \cite{Ga,Fe1,Fe2,Fe3,Fe4}.
In our case here, for $\omega_\mu(t)=(\omega_0(t),0,0,0)$
consistent with a FRW metric, if $\omega_0(t)^2\sim 1/\phi(t)$
then a similar solution $\phi(t)\ra$constant can exist.
Assuming that  $\phi$  acquires a vev by such  mechanism or at the  quantum level, etc,
then equation (\ref{jj})  gives  $\nabla_\mu \w^\mu=0$.
This is  the ``gauge fixing'' condition, specific  to a massive Proca field, that emerges
from the conserved current of the Weyl gauge symmetry.

Finally, one may ask what Weyl geometry tells us about the  cosmological constant ($\Lambda$).
From Lagrangians (\ref{alt2}) and (\ref{EP}) we find
\bea\label{la}
\Lambda=\frac14 \,\langle\phi^2\rangle, \qquad \frac{\Lambda}{M_p^2}=\frac 32 \,\xi^2.
\eea
Both $\Lambda$ and $M_p$ are
generated by same $\phi$ and are thus related and  $\Lambda> 0$. For  $M_p$ fixed, $\Lambda$ is
small because gravity is weak ($\xi\ll 1$).
In the limit $\langle\phi\rangle\ra 0$ then $\Lambda, M_p\ra 0$ and the Weyl
gauge symmetry is restored\footnote{This limit is formal, since the linearisation of (\ref{inA})
with $\phi^2=-\tilde R$ implicitly  assumes that $\phi$ is non-zero.}.
This shows how $\Lambda$ is protected by this  symmetry.

In conclusion,  Weyl action (\ref{inA}), (\ref{ST}) is
more fundamental than Einstein-Proca action (\ref{EP}) which is its ``low-energy'', broken phase.
When the massive Weyl gauge boson decouples, the  geometry becomes Riemannian and
the Einstein gravity is recovered. In a sense this picture is entirely {\it geometrical}
\cite{Ghilencea:2022lcl}  since we did not yet include any matter.
Thus, ultimately the underlying geometry of our Universe may actually be  Weyl conformal geometry.
Its Weyl gauge symmetry could then  explain a small (non-vanishing) positive
cosmological constant.

\subsection{Weyl quadratic gravity and  ``photon'' -  photon mixing}\label{s2.2}

Consider now $\cL_0$ in the presence of the SM hypercharge gauge group $U(1)_Y$.
A kinetic mixing  of $\w_\mu$ (Weyl ``photon'') with the $B_\mu$ gauge field of  U(1)$_Y$
is allowed by the direct product symmetry $U(1)_Y\times D(1)$. 
Such mixing was mentioned in the literature  \cite{Guendelman} but not investigated.
Consider then
\medskip
  \bea\label{L1}
\cL_1=\sg\,\Big\{\,
\frac{1}{4!\,\xi^2}\,\tilde R^2
-\frac{1}{4} \, \Big[\, F_{\mu\nu}^{\,2}+ 2\sin \chi\, F_{\mu\nu}\, F_y^{\mu\nu}
+ F_{y\,\mu\nu}^{\,2}\Big] -\frac{1}{\eta^2} \tilde C_{\mu\nu\rho\sigma}^2\Big\}.
\eea

 \medskip\noindent
where $F_y$ is the field strength of $B_\mu$.
The source of  $B_\mu$ is 
the SM fermionic Lagrangian (not shown in (\ref{L1})) which is 
invariant under (\ref{WGS}) and does not depend on $\w_\mu$ \cite{Kugo}
(see next section). This  mixing could eventually be forbidden by some
discrete symmetry, not discussed here\footnote{
  Note that $\w_\mu$ is C-even \cite{Kugo} and  the photon is $C$-odd and the mixing 
  violates C
  and CP.
  Global or discrete symmetries (C, CP,  $Z_2$ etc) can be used to forbid  the
  kinetic  mixing; such symmetries
  can however be  broken by black-hole physics/gravity \cite{Kallosh}. Also, the  CPT invariance theorem 
  applies only if the theory is  local,  unitary and in flat  space-time, so it cannot be used here:
  the
  Weyl-geometry  actions are neither unitary ($\tilde C^2$ term in Weyl action
  has a ghost) nor in flat space-time. The consequences of  $\chi\not=0$
  are further studied in Section~\ref{s2.7}.}.

We repeat the steps  in Section~\ref{s2.1} and after
transformation (\ref{WGS}) under which $B_\mu$ is invariant,
$\hat B_\mu=B_\mu$, we find $\cL_1$ in terms of the  new fields (with a hat):
\medskip
\be
\label{EPp}
\cL_1\!=\!\sgh \Big\{- \frac12 M_p^2\hat R +\frac34 M_p^2\,\q^2\hat\w_\mu \hat \w^\mu
- \frac{3 \xi^2}{2}\,M_p^4
-\frac{1}{4} \, \Big[\,\frac{1}{\gamma^2} \hat F_{\mu\nu}^{\,2}
+ 2\sin \chi\, \hat F_{\mu\nu}\, \hat F_y^{\mu\nu}
+ \hat F_{y\,\mu\nu}^{\,2}\Big]
-\frac{1}{\eta^2} C_{\mu\nu\rho\sigma}^2\Big\}.
\ee

\medskip
The kinetic mixing is removed by a  transformation \cite{JMR} to  new ('primed') fields
 \medskip
\bea\label{bwp}
\hat \w_\mu  = \gamma\, \w'_\mu \sec\tilde\chi,
% \nonumber\\
\qquad
\hat B_\mu =  B'_\mu-\w'_\mu\,\tan\tilde\chi,
\qquad\text{with}\quad \sin\tilde\chi\equiv\gamma\sin\chi,\,\,
\eea

\medskip\noindent
where, for a simpler notation, we introduced $\sin\tilde\chi$
(note that $\gamma\leq 1$)\footnote{In the limit
  $\gamma=1$ there is no  $\tilde C_{\mu\nu\rho\sigma}^2$ term in the initial
  action (formally $\eta\ra \infty$).}.
The result is
\medskip
\be\label{LLp}
\cL_1=\sgh\,\, \Big\{
-\frac12 \,M_p^2 \hat R +\frac{1}{\eta^2} C_{\mu\nu\rho\sigma}^2
+\frac34 \,M_p^2\,\q^2\,
\gamma^2\sec\tilde\chi^2\,\,\w'_\mu{\w'}^{\mu} 
-\frac14\,(  F_{\mu\nu}^{\prime\, 2} \,+\ F_{y\,\mu\nu}^{\prime\, 2})
-\frac{1}{\eta^2} \,C_{\mu\nu\rho\sigma}^2\Big\},
\ee

\medskip\noindent
with $ F'_{y\,\mu\nu}=\nabla_\mu B'_\nu -\nabla_\nu B'_\mu=
\partial_\mu B'_\nu -\partial_\nu B'_\mu$ and
$F'_{\mu\nu}=\partial_\mu\w'_\nu-\partial_\nu\w'_\mu$.

As in the previous section, we obtain again the Einstein-Proca action but with
diagonal gauge kinetic terms for both gauge fields.
However, the final, canonical hypercharge gauge field $B_\mu'$ has acquired
a dependence on the  Weyl gauge field, see (\ref{bwp}), due to the initial kinetic mixing.
In the full model, upon the electroweak symmetry breaking the photon field ($A_\mu$)
is a mixing of the hypercharge ($B'_\mu$) with SU(2)$_L$ neutral gauge field ($A_\mu^3$)
\medskip
\be
A_\mu=B'_\mu \cos\theta_w +A_\mu^3\,\sin\theta_w 
=\big[\hat B_\mu +\hat \w_\mu \sin\chi\big]\,\cos\theta_w
+\sin\theta_w A_\mu^3.
\ee

\medskip\noindent
where $\theta_w$ is the Weinberg angle and in the second step  we used eq.(\ref{bwp}).

Due to the gauge kinetic mixing the
photon field includes a  small component of the initial Weyl gauge field,
suppressed by  $\sin\chi$ and by the mass ($\sim M_p$) of $\w_\mu$, but still
present\footnote{In some sense this says that  Weyl's unfortunate attempt
  to identify $\w_\mu$ to the photon was not entirely wrong, if the aforementioned mixing
  is present.};
however, {\it it
exists only in the presence of matter e.g. fermionic fields}
that act as the source of $B_\mu$.
Such mixing in models with Abelian gauge fields beyond  the hypercharge exists
in string models,  with similar massive {\it and} anomaly-free
gauge fields (as $\w_\mu$, see later) and  similar mass mechanism \cite{Quevedo}.
However,  here $\w_\mu$ is  a gauge field of a space-time (dilatation) symmetry.
The mixing is not forbidden by the Coleman-Mandula theorem - the overall
symmetry is always a direct product $U(1)_Y\times D(1)$ and
both  symmetries are subsequently broken spontaneously
\footnote{{The theorem   implies that  $D(1)$ cannot be part of an
  internal non-Abelian symmetry so $d$ cannot be fixed}.}.

\subsection{ Fermions}\label{s2.3}

Consider now the SM fermions ($\psi$)   in Weyl  geometry and examine their action.
To begin with, to avoid
a complicated notation we do not display the SM  gauge group dependence:
\medskip
\bea\label{inLf}
\cL_f=\frac12\,\sg\,\, \overline\psi \, i\,\gamma^a \, e^\mu_{\,a}\,\tilde\nabla_\mu \psi+h.c.,
\qquad
\tilde\nabla_\mu\psi=
\Big(\partial_\mu  -  \frac34\, \alpha\ \hat \w_\mu +
\frac12\,\tilde s_\mu^{\,\,ab}\,\sigma_{ab}\Big)\psi.
\eea

\medskip\noindent
Here $\tilde s_\mu^{\,\,ab}$ is the Weyl geometry spin connection.
In (\ref{inLf}),  the Weyl charge of the fermions is $(-3/4)$ according to our convention
in (\ref{WGS}) ($d=1$). The relation of the  Weyl spin connection  to the 
spin connection $s_\mu^{\,\,ab}$  of (pseudo-)Riemannian geometry is (see Appendix~\ref{AA})
\medskip
\bea
\tilde s_\mu^{\,\,ab}
& =&
s_\mu^{\,\,ab}+\frac12 \,\alpha\, (e_\mu^a \,e^{\nu b}-e_\mu^b e^{\nu a})\,\,\hat\w_\nu,
\nonumber\\[7pt]
s_\mu^{\,\,ab}& =&  (-1)\, e^{\lambda b}\, (\partial_\mu\,  e_\lambda^{\,\,a} - e_\nu^{\,\,a}\,
\Gamma_{\mu\lambda}^\nu ),
\eea

\medskip\noindent
where $\sigma_{ab}=\frac14 [\gamma_a,\gamma_b]$ while
$\Gamma_{\mu\lambda}^\nu$ is the Levi-Civita connection,
$g_{\mu\nu}=e_\mu^{\,\,a} \,e_{\nu}^{\,\,b} \eta_{ab}$ and $e^\mu_{\,\,a} e_\nu^{\,\,a}=\delta^\mu_\nu$.
It can be checked that, similar to the Weyl connection ($\tilde\Gamma$),
the Weyl spin connection $\tilde s_\mu^{\,\,ab}$ is invariant under (\ref{WGS}). This is seen by
using that $s_\mu^{\,\,ab}$ transforms under (\ref{WGS})  as
%\medskip
\bea
\hat s_\mu^{\,\,ab}=s_\mu^{\,\,ab} +
(e_\mu^a \,e^{\nu b} -e_\mu^b \,e^{\nu a})\,\partial_\nu\ln\Sigma^{1/2}. % \Omega
\eea

\medskip\noindent
With $\tilde s_\mu^{\,\,ab}$ invariant, one checks that $\cL_f$ is Weyl gauge invariant.
In fact one can easily show that $(-3/4) \alpha\, \hat\w_\mu$ in 
$\gamma^\mu\,\tilde\nabla_\mu\psi$
is cancelled by the $\hat\w_\mu$-presence in the Weyl spin connection.
This cancellation also
happens between fermions and anti-fermions
\cite{Kugo} (eqs. 36, 37)\footnote{but only for the Weyl charge in (\ref{inLf}) can we write a
  Weyl invariant $\cL_f$ without a\,scalar compensator in~\cite{Kugo}.}.
This is so  because both fermions and anti-fermions have the same
{\it real} Weyl charge (no $i$ factor in $\tilde\nabla_\mu\psi$).
As a result, we have
\bea
\cL_f=\frac12\,\sg\,\, \overline\psi \, i\,\gamma^a \, e^\mu_{\,a}\,\nabla_\mu \psi+h.c.,
\qquad \nabla_\mu \psi=\Big(\partial_\mu +\frac12  \, s_\mu^{\,\,ab} \, \sigma_{ab}\Big)\,\psi.
\eea

\medskip\noindent
Thus the SM fermions  do not couple \cite{Kugo} to the Weyl field $\w_\mu$ 
and there is no gauge anomaly.

We can now restore the SM gauge group dependence and the Lagrangian becomes
\medskip
\bea\label{Lf}
\cL_f & =&\frac12\, \sg\,\, \overline\psi \, i\,\gamma^a \, e^\mu_{\,a}\,
\,\Big[
\partial_\mu - i g\, \vec T \vec{\hat A}_\mu - i\, Y g^\prime \hat B_\mu
+\frac12 s_\mu^{\,\,ab} \,\sigma_{ab} \Big]\,\psi+h.c.,
\eea

\medskip\noindent
with the usual quantum numbers of the fermions under the SM  group (not shown), $\vec T=\vec\sigma/2$,
and with $g$ and $g^\prime$ the gauge couplings of $SU(2)_L$ and $U(1)_Y$.
But this is not the final result.

Since the fermions are $U(1)_Y$ charged and the initial % hypercharge
field $\hat B_\mu$ in (\ref{Lf})  is shifted   by the gauge kinetic mixing, as seen in eq.(\ref{bwp}),
then  $\w'_\mu$ is still present in $\cL_f$:
\medskip
\bea\label{Lff}
\cL_f & =& \frac12\, \sg\,\, \overline\psi \, i\,\gamma^a \, e^\mu_{\,a}\,
\,\Big[
\partial_\mu - i g\, \vec T \vec{\hat A}_\mu - i\, Y g^\prime\,
\big(B'_\mu-\w'_\mu\tan\tilde\chi\big)
+\frac12 s_\mu^{\,\, ab} \,\sigma_{ab} \Big]\,\psi+h.c.
\eea

\medskip
We found a new coupling of  the SM fermions to $\w_\mu^\prime$, of strength $Y g^\prime \tan\tilde\chi$.
This coupling comes  with the usual fermions hypercharge assignment (which is anomaly-free).
After the electroweak
symmetry breaking  $B'_\mu$ is replaced in terms of the mass eigenstates $A_\mu$, $Z_\mu$, $Z^\w_\mu$
and $\w'_\mu$ is a combination of $Z_\mu$, $Z_\mu^\w$  (see later,  eq.(\ref{U})).
If  $\chi\!\sim\!\tilde\chi\!=\!0$, the fermions Lagrangian is identical to that in
the  (pseudo-)Riemannian case  (with  no Weyl gauge symmetry).

Regarding the Yukawa interactions notice that the SM  Lagrangian is\,invariant under (\ref{WGS})
\smallskip
\bea\label{Y}
\cL_Y=\sqrt{g} \sum_{\psi=l,q} \Big[\overline \psi_{L} Y_\psi H \psi_{R}
+ \overline \psi_{L} Y'_\psi \tilde H \psi_{R}'\Big] + h.c.
\eea
where $H$ is the Higgs $SU(2)_L$ doublet and $\tilde H=i\sigma_2 H^\dagger$,
the sum is over leptons and quarks; $Y, Y'$ are the SM  Yukawa matrices.
$\cL_Y$   is invariant under (\ref{WGS}): indeed,
since the Weyl charge is real,
the sum of charges of the fields in each Yukawa term is  vanishing:
two fermions (charge $2\times (-3)/4$),
 the Higgs (charge $-1/2$) and $\sqrt{g}$ (charge $2$). Hence the Yukawa interactions have the same form
as in SM in the (pseudo-)Riemannian space-time.

\subsection{Gauge bosons}\label{s2.4}

Regarding the SM gauge bosons,  their SM action is
 invariant under transformation (\ref{WGS}) \cite{Kugo}.
A  way to understand this is that a gauge boson of the SM enters under the
corresponding covariant derivative acting on a field charged under it
and should transform (have same weight) as  $\partial_\mu$ acting on that field;
 since coordinates are kept fixed under (\ref{WGS}), the gauge fields do not transform either.
Their kinetic terms are then  similar  to those of the SM in  flat space-time, since
the Weyl  connection is symmetric. Explicitly, this is seen  from the equation below, where
the sum is over the SM gauge group factors:  $SU(3)\times SU(2)_L\times U(1)_Y$
\bea
\cL_g=-\sum_{\rm{groups}}\frac{\sg}{4} g^{\mu\rho} g^{\nu\sigma} F_{\mu\nu} F_{\rho\sigma},
\eea
%\medskip\noindent
 $F_{\mu\nu}$ involves the difference $\tilde\nabla_\mu A_\nu -\tilde\nabla_\nu A_\mu$, where
$A$ is a generic notation for a SM gauge boson and
since $\tilde\nabla_\mu A_\nu=\partial_\mu A_\nu -\tilde\Gamma_{\mu\nu}^\rho A_\rho$, then for a symmetric
$\tilde\Gamma_{\mu\nu}^\rho=\tilde\Gamma_{\nu\mu}^\rho$ one sees that $\tilde\Gamma$ and its 
$\w_\mu$-dependence cancel out in the field strength $F_{\mu\nu}$.
Hence, $\cL_g$ does not depend on $\w_\mu$ and has the same form in Weyl and 
in (pseudo)Riemannian geometries.

\subsection{Higgs sector}\label{s2.5}

\noindent
{$\bullet$\bf\, The action:}
Let us now consider the SM Higgs  doublet ($H$)  in Weyl conformal geometry:
  \medskip
  \be\label{higgsR}
\cL_H\!=\!\sg\,\Big\{\,
\frac{\tilde R^2}{4!\,\,\xi^2}
-\frac{\tilde C_{\mu\nu\rho\sigma}^2}{\eta^2}
\,-\,\frac{\xi_h}{6}\,\vert H \vert^2 \tilde R +\vert \tilde D_\mu H\,\vert^2
-{\lambda}\, \vert H\vert^4 %\Big)
% \nonumber \\[5pt]
% &-&
-\frac{1}{4}  \Big( F_{\mu\nu}^{\,2}+ 2\sin \chi\, F_{\mu\nu}\, F_y^{\mu\nu}
+ F_{y\,\mu\nu}^{\,2}\Big)\Big\}.
\ee

\medskip\noindent
The $SU(2)_L \times U(1)_Y \times D(1)$ derivative acting on $H$ is
% \medskip
\bea\label{deriv}
\tilde D_\mu H&=&\big[\partial_\mu - i \cA_\mu - (1/2)\, \q\,\w_\mu\big]\, H,
%\qquad
\eea

\smallskip\noindent
where $\cA_\mu=(g/2) \,\vec \sigma.\vec A_\mu +  (g^\prime/2)\, B_\mu$;\,
$\vec A_\mu$ is the $SU(2)_L$ gauge boson,
$B_\mu$ is the $U(1)_Y$ boson.
The case  of no gauge kinetic mixing in (\ref{higgsR}) ($\chi\!=\!0$) is obvious. We keep
$\chi\!\not=\!0$ for generality.

We consider the electroweak  unitary gauge where
$H=(1/\sqrt{2})\, h \, \zeta$, with $\zeta^T\!\equiv\!(0,1)$. Then 
\bea
\vert \tilde D_\mu H\vert^2&=&\vert (\partial_\mu - \q/2\,\, \w_\mu) H\vert^2
+H^\dagger \cA_\mu \cA^\mu H,
\eea
with
\bea
H^\dagger \cA_\mu\cA^\mu H=(h^2/8)\,\cZ,\qquad 
\cZ\equiv \big[ g^2 (A_\mu^{1\,2}+A_\mu^{2\,2})+(g A_\mu^3-g^\prime B_\mu)^2\big].
\eea

\medskip\noindent
As done earlier,  in  $\cL_H$ replace
$\tilde R^2\ra -2\phi^2 \tilde R-\phi^4$ to find a classically equivalent
action; using the equation of motion of $\phi$ and its solution $\phi^2=-\tilde R$
back in the action, one recovers (\ref{higgsR}). After this replacement,
the  non-minimal coupling term in (\ref{higgsR}) is modified
 \smallskip
\bea\label{rep}
-\frac{1}{6}\,\xi_h\,\vert H\vert^2
\,\tilde R \ra \frac{-1}{12} \,\Big(\,\frac{1}{\xi^2} \phi^2+\,\xi_h\,h^2\Big)\,\tilde R.
\eea

\medskip\noindent
It is interesting to notice that the initial term  in the action,
$(1/\xi^2)\,\tilde R^2$, (where $\xi\!<\!1$)  in (\ref{higgsR})  was replaced
by  a term above with a large  non-minimal coupling  $1/\xi^2\!>\!1$ (plus an additional $\phi^4$).
For details, the
full Lagrangian $\cL_H$ after step (\ref{rep}) is shown in the Appendix,  eq.(\ref{action2}).

Next,  to fix the gauge,  apply transformation (\ref{WGS}) to $\cL_H$  with  a special
{\it scale-dependent} $\Sigma$  which fixes the fields combination $(\phi^2/\xi^2+\xi_h h^2)$
to a constant:
\medskip
\bea %%%%   \Sigma=\Omega^2 
\hat g_{\mu\nu}\!=\! \Sigma\,g_{\mu\nu},\,\,\,\,
\hat\phi^2=\!\frac{\phi^2}{\Sigma},\,\,\,\,\,
\hat \w_\mu\!=\!\w_\mu\! -\!\frac1\q \partial_\mu\ln\Sigma,
\,\,\,\,
\hat B_\mu\!=\! B_\mu,
\,\,\,\,
\hat\cA_\mu\!=\!\cA_\mu,
\,\,\,\,
\Sigma\!\equiv 
\!\frac{\phi^2/\xi^2+\xi_h h^2}{\big\langle\phi^2/\xi^2+\xi_h\,h^2\big\rangle}.
\eea

\medskip\noindent
In terms of the transformed fields and metric (with a `hat'), $\cL_H$ becomes
\medskip
\bea
\cL_H\!\!\!&=&\!\!\!\!\!\sgh\,\,\Big\{
-\frac12 M_p^2 \Big[ \hat R  - 3 \q \nabla_\mu \hat \w^\mu-\frac32 \q^2\hat\w_\mu \hat \w^\mu \Big]
- \frac{1}{\eta^2} C_{\mu\nu\rho\sigma}^2
+\frac12\, \big\vert (\partial_\mu -\q/2\, \hat \w_\mu)\,\hat h\big\vert^2
+\frac18\, \hat h^2\, \hat\cZ 
\nonumber\\[2pt]
&-&\hat V - \frac{1}{4} \, \Big(\frac{1}{\gamma^2} \hat F_{\mu\nu}^{\,2}
+ 2\sin \chi\, \hat F_{\mu\nu}\, \hat F_y^{\mu\nu}
+ \hat F_{y\,\mu\nu}^{\,2}\Big)\Big\}, 
\label{iL}\eea

\noindent
were we used (\ref{def}), the notation
$\hat \cZ=\cZ(B_\mu\ra\hat B_\mu,\vec A_\mu\ra \hat{\vec{A_\mu}})$, with $\gamma\leq 1$ defined in
(\ref{ST}) and
\smallskip
\be\label{planck}
M_p^2\equiv \frac16 \,\Big\{\frac{1}{\xi^2}\,\big\langle\phi^2\big\rangle +\xi_h \,\big\langle
h^2\big\rangle\Big\},\qquad
\ee
and finally
\be
\label{neg}
\hat V=\frac{1}{4!}\,\Big[ 6\,\lambda \, \hat h^4+ \xi^2\,(6 M_p^2 - \xi_h \, \hat h^2)^2\Big].
\ee

\bigskip\noindent
We found again a massive $\w_\mu$ in (\ref{iL}) by  Stueckelberg mechanism after `eating'
the {\it radial direction} field $(1/\xi^2\, \phi^2+\xi_h h^2)$,
with constraint $\nabla_\mu \w^\mu=0$.
We identify $M_p$ with the Planck scale; $M_p$ and thus also $m_\w$ 
receive now contributions from both the Higgs and $\phi$.

The term proportional to $\xi^2$ in $\hat V$ is ultimately due to the $(1/\xi^2)\,\tilde R^2$ term
in the action and is ultimately responsible for the EW symmetry
breaking and for inflation, see later.

Eq.(\ref{iL}) contains a mixing term $\hat \w^\mu \partial_\mu \hat h$
from the Weyl-covariant derivative of $\hat h$. We
choose  the  unitary gauge for the D(1) symmetry i.e. eliminate this term
by replacing 
\medskip
\bea\label{hw}
\hat h&=&M_p\sqrt{6} \sinh\frac{\sigma}{M_p\sqrt 6},%\nonumber
%\\
\qquad\qquad
\hat\w_\mu= \hat\w_\mu'+\frac{1}{\q} \partial_\mu\ln\cosh^2 \frac{\sigma}{M_p\sqrt{6}}.
\eea

\noindent
Then $\cL_H$ becomes 
\medskip
\bea
\cL_H&=& \sgh\,\Big\{
-\frac12 M_p^2\,\hat R - \frac{1}{\eta^2} C_{\mu\nu\rho\sigma}^2+
\frac34  M_p^2 \q^2\, \hat\w_\mu^\prime\hat \w^{\prime \mu}
\cosh^2 \frac{\sigma}{M_p\sqrt{6}} 
+
\frac12 (\partial_\mu\sigma)^2-\hat V
\nonumber\\[4pt]
&+& 
\frac34\, M_p^2\, \hat\cZ\,\sinh^2\frac{\sigma}{M_p\sqrt{6}}
- \frac{1}{4} \, \Big(\frac{1}{\gamma^2} \hat F_{\mu\nu}^{\prime\,2}
+ 2\sin \chi\, \hat F^\prime_{\mu\nu}\, \hat F_y^{\mu\nu}
+ \hat F_{y\,\mu\nu}^{\,2}\Big)\Big\}.
\eea

\medskip\noindent
with the potential $\hat V$ expressed now in terms of the actual Higgs
field $\sigma$, using (\ref{neg}), (\ref{hw}).

The term $(3/4) \, M_p^2\,\q^2 \,\hat\w'_\mu\,\hat\w^{\prime\mu} \cosh^2[\sigma/(M_p\sqrt{6})]$,
after expanding it for $\sigma\!\leq\! M_p$, 
contains a leading coupling $(\q^2/8)\,\hat\w'_\mu\hat\w^{\prime \,\mu}\sigma^2$,
plus additional corrections suppressed by $M_p^2$.
{\it If there is no kinetic mixing, $\chi=0$, this is the only coupling of $\w_\mu$ to the Higgs 
and to the SM states}!

\medskip
\noindent
{$\bullet$ \bf Kinetic mixing:}
Finally,  remove the gauge  kinetic mixing in $\cL_H$ by replacing $\hat\w_\mu'$, $\hat B_\mu$ by
%\medskip
\bea\label{bw}
\hat \w_\mu' & = &\gamma\, \w'_\mu \sec\tilde\chi,
\qquad % \nonumber\\[1pt]
 \hat B_\mu= B'_\mu-\w'_\mu\,\tan\tilde\chi,
 \qquad (\sin\tilde\chi\equiv\gamma\sin\chi);\,\,
\eea

\medskip\noindent
and  $\cL_H$ becomes:
\bea\label{LL}
\cL_H&=&\sgh\,\, \Big\{
-\frac12 \,M_p^2 \hat R
-\frac{1}{\eta^2} C_{\mu\nu\rho\sigma}^2
+\frac34 \,M_p^2\,\q^2\,
\gamma^2(\sec^2\tilde\chi)\,\,\w'_\mu{\w'}^\mu %
+ \frac12 (\partial_\mu\sigma)^2 -\hat V
\nonumber\\[5pt]
&+&\frac34\,M_p^2\,
\,\Big[\cZ' +\q^2 \gamma^2\,(\sec^2\tilde\chi)\,
\w'_\mu {\w'}^\mu\Big]\,\sinh^2 \frac{\sigma}{M_p\sqrt 6}-
\frac14\,( F_{\mu\nu}^{\prime\,2} \,+ F_{y\,\mu\nu}^{\prime\,2})\Big\}.
\eea
where $F'$ ($F'_y$) is the field strength of $\w'$ ($B^\prime$)  and
\bea
\cZ'=
\Big[ g^\prime (B'_\mu -\w'_\mu \tan\tilde\chi)
-g \,\hat A_\mu^3\Big]^2 + g^2 (\hat A_\mu^{1\,2}+\hat A_\mu^{2\,2})
\eea

\medskip\noindent
Note the presence in $\cL_H$  of a coupling 
$\Delta\cL_H= (1/8)\,\,\sigma^2\, \,\w'_\mu\w^{\prime \mu} \,(g^{\prime 2}
\tan^2\tilde\chi+\alpha^2 \,\gamma^2\sec^2\tilde\chi)$. This
 is due to 1)  the gauge  kinetic mixing $\chi$ and 2)  to the
 Higgs coupling  to $\w_\mu$,  eq.(\ref{deriv}).
 This coupling  is non-zero
 {\it  even if there is no gauge kinetic mixing} ($\chi\!=\!0$), when it becomes
 \bea
 \Delta\cL_H= (1/8) \alpha^2\gamma^2\,\sigma^2\w'_\mu\w^{\prime \mu}.
 \eea
 This is relevant for Higgs physics  and can constrain $\alpha$.
 If $\gamma=1$ (i.e. if there is no $\tilde C_{\mu\nu\rho\sigma}^2$ term in (\ref{higgsR}))
 then this coupling is due entirely to the higgs kinetic term $\vert D_\mu H\vert^2$ in (\ref{higgsR})
 and is the only coupling of the  SM to the  background Weyl geometry (apart from that to the graviton).
 In the symmetric phase $\Delta \cL_H$ can generate higgs production via
Weyl boson fusion. This coupling is further discussed in \cite{Ghilencea:2022lcl}.

\medskip\noindent
{$\bullet$ \bf Higgs potential:}
\,\,\,\, One may write  $\cL_H$ in a more compact form
\medskip
\be\label{massmatrix}
\cL_H=\sgh
\,\, \Big\{
\frac{-1}{2} \,M_p^2 \hat R
- \frac{1}{\eta^2} C_{\mu\nu\rho\sigma}^2
-\frac14\, ( F_{\mu\nu}^{\prime 2} +F_{y\, \mu\nu}^{\prime 2})
+\cL_h
+ m_W^2(\sigma) W^-_\mu W^{+ \mu} 
+
\frac12 \, X^T \cM^2(\sigma) X
\Big\}
\ee
%\medskip\noindent
with the $\sigma$-dependent mass  $m_W(\sigma)$ of  SU(2)$_L$
bosons $W_\mu^\pm=1/\sqrt 2 \,(A_\mu^1 \mp i A_\mu^2)$ given by
%\medskip
\be
m_W^2(\sigma)=\frac{3 g^2}{2} M_p^2 \,\sinh^2\frac{\sigma}{M_p\sqrt 6}
=\frac{g^2}{4}\,\sigma^2+ \cO(\sigma^4/M_p^2).
\ee
%\medskip\noindent
The  $\sigma$-dependent matrix $\cM(\sigma)$
written in eq.(\ref{massmatrix}) in the
basis   $X\!\equiv\!(B'_\mu, A_\mu^3,  \w'_\mu)$
is presented  in the Appendix, eq.(\ref{mass1}). Finally
we have
\bea
\cL_h=\frac12 (\partial_\mu \sigma)^2- \hat V(\sigma)
\eea
\bea\label{Vinflation}
\hat V(\sigma)
&=&
\frac{3}{2} \,M_p^4\, 
\Big\{{6 \lambda}\, \sinh^4 \frac{\sigma}{M_p\sqrt 6}
+
\xi^2\Big(1- \xi_h \sinh^2\frac{\sigma}{M_p\sqrt 6}\Big)^2\Big\}
\\[6pt]
&=&
\label{smallfield}
\frac{1}{4} \,\Big(\lambda - \frac19 \,\xi_h\,\xi^2 +\frac16\xi_h^2\,\xi^2 \Big)\,\sigma^4
-
\frac12\,\xi_h\xi^2 M_p^2 \,\sigma^2 %
+
\frac{3}{2}\,\xi^2 M_p^4+\cO(\sigma^6/M_p^2).
\eea
%\medskip
This is the Higgs potential in our SMW model in the unitary gauge for
 the  EW and $D(1)$ symmetries.
 The second line is valid  for small field values   $\sigma\ll M_p$ when
 we recover  a  Higgs potential similar to that in the SM;
 the quadratic term has a negative coefficient (with $\xi_h>0$, as needed for inflation,
 see later). This follows when the  Higgs field contributes positively
to the Planck scale, eq.(\ref{planck}) and ``to compensate'' for its  contribution
to $M_p$, a negative sign emerges in (\ref{neg})  and in $\hat V(\sigma)$.
The EW  symmetry is thus broken  at tree level.

\subsection{EW scale and Higgs mass}\label{s2.6}

The small field regime $\sigma\!\ll\! M_p$ in (\ref{smallfield})
gives realistic predictions  in the limit $\xi_h\,\xi^2\!\ll\! 1$; indeed, 
in this case the quartic Higgs coupling becomes $\lambda$ and the
EW scale $\langle\sigma\rangle$ and Higgs mass are
\bea\label{min}
\langle\sigma\rangle^2=\frac{1}{\lambda}\,\xi_h\,\xi^2\,M_p^2,
\qquad
m_\sigma^2=2\,\xi_h\,\xi^2\,M_p^2.
\eea

To comply with the values of the Higgs mass and EW vev
we must set  $\xi\,\sqrt{\xi_h}\sim 3.5 \times 10^{-17}$. 
This  means one or both perturbative couplings $\xi_h$ and $\xi$  take
small values, while  $\lambda\sim 0.12$ as in the SM and
the regime $\sigma\ll M_p$ is  respected.
Recall that  $\xi$ is the coupling of the term  $(1/\xi^2)\tilde R^2$ in the action,
hence we see the importance of this term for  the hierarchy of scales!

From (\ref{min}),  using the Planck scale expression eq.(\ref{planck}), then
% \medskip
\bea\label{vevphi0}
\langle\sigma\rangle^2 \approx \frac{\xi_h}{6\lambda}\,\langle\phi^2\rangle.
\eea

With $\xi \sqrt\xi_h$ fixed earlier, one still has a freedom of
either a hierarchy or  comparable values  of these two  vev's, depending on
the exact  values of  $\xi_h<1$.
Eq.(\ref{vevphi0})  relates the EW  scale physics to
 the underlying Weyl geometry represented  by the $\tilde R^2$  term in the action
 (from which $\phi$ is ``extracted'').

The  SMW model  with the
Higgs action as in eqs.(\ref{higgsR}), (\ref{massmatrix}) has   similarities  to
Agravity \cite{Strumia1,Strumia2}  which is a global scale invariant model.
Unlike in Agravity,  we only have the Higgs scalar, 
while  the role of the second scalar field ($s$) in \cite{Strumia1},
that generated the Planck scale and Higgs mass in Agravity is played in our model
by the ``geometric'' Stueckelberg field ($\phi$); \,\, $\phi$ was not added ``ad-hoc''
and cannot couple to the Higgs field, being extracted from the  $\tilde R^2$ term
itself (see eq.(\ref{higgsR})).
Hence, there is no classical coupling between the Higgs field and the field generating $M_p$,
while in \cite{Strumia1} a coupling $\lambda_{HS} h^2 s^2$ is present.

However, the SMW  contains the field  $\w_\mu$ (part of Weyl geometry),
not present in \cite{Strumia1}.
Our preference here for a local, gauged scale symmetry, that brought in the Weyl gauge field,
is motivated by three aspects:  firstly, we already have a ``geometric'' mass generation
mechanism which does not need adding ad-hoc an extra scalar; secondly,
global symmetries do not survive black-hole physics \cite{Kallosh}
and finally, the  Weyl gauge symmetry of the action is also a symmetry of
the underlying  geometry (connection $\tilde\Gamma$), as it should
be the case.

At the quantum level,  large loop  corrections to  $m_\sigma$ could in principle
arise, as in the SM (plus those  due to  $\w_\mu$ by the coupling $\w_\mu w^\mu \sigma^2$).
But the Weyl gauge symmetry can change this.
The  mass $m_\w\!\sim\! \q M_p$
may be light (if   $\q\ll 1$), possibly not far above the lower
bound (of few TeV)  on the non-metricity scale (set by  $m_\w$) \cite{Latorre}.  
This means the  Weyl gauge symmetry breaking scale can be low.
The mass of $\w_\mu$  is then the highest physical
scale (``cutoff'') for the low-energy observer. Then all  quantum corrections
to $m_\sigma^2$ are expected to be quadratic in the scale of ``new physics''
($m_\w^2$), so $\delta m_\sigma^2\propto m_\omega^2$.
Above $m_\w$ the  gauged scale symmetry is restored, 
together with its UV protection (for the Higgs mass) not affected by its
spontaneous breaking. In this way the Weyl gauge symmetry (with  $\xi,\q\!\ll\! 1$)
could  give a solution to the hierarchy problem. Intriguingly,
since $m_\w$ also sets the non-metricity scale, this suggests
the  {\it hierarchy problem and non-metricity scale are related!}

\subsection{Constraints from  $Z$ mass}\label{s2.7}

Let us now compute  the eigenvalues of the
Higgs-dependent matrix $\cM^2(\sigma)$, eqs.(\ref{massmatrix}),
(\ref{mass1}), and examine the constraints from the mass of $Z$
on the model  parameters   $\q$ and  $\chi$.
Since $Z_\mu$ and $\w_\mu$ mix, part of $Z$ boson mass is not
due the Higgs mechanism, but to this mixing and ultimately,
to the Stueckelberg mechanism giving mass to $\w_\mu$. 
After the electroweak  symmetry breaking, 
in the mass eigenstates basis of  $\cM^2(\langle\sigma\rangle)$, one has
the photon field  ($A_\mu$) (it is massless, since $\det \cM^2=0$),
the neutral gauge boson ($Z$) and  the  
Weyl field ($Z^\w$).

$\cM^2(\sigma)$ is brought to diagonal form by two rotations
(\ref{mix1}), (\ref{mix2}) giving
%
%\medskip
\bea
\begin{pmatrix}\label{U}
  B'_\mu\\
  A_\mu^3\\
  \w'_\mu
\end{pmatrix}
=\begin{pmatrix}
  \cos\theta_w  & -\sin\theta_w \cos\zeta & -\sin\theta_w \sin\zeta\\
   \sin\theta_w & \cos\theta_w\cos\zeta  & \cos\theta_w\sin\zeta\\
  0  & -\sin\zeta & \cos\zeta
 \end{pmatrix}
\begin{pmatrix}
  A_\mu\\
  Z_{\mu}\\
  Z^\w_{\mu}
\end{pmatrix}
\eea

\medskip\noindent
Denote  by $U$ the matrix relating the gauge eigenstates  $(B'_\mu$, $A_\mu^3$, $\w'_\mu)$
 to the  mass  eigenstates $(A_\mu$, $Z_\mu$, $Z_\mu^\w)$;  then $\cM^2(\sigma)$ is diagonalised into
$\cM_d^2=U^T \cM^2 U$ for a  suitable $\zeta$
\medskip
\bea\label{ze}
\tan 2 \zeta = \frac{-2 g^\prime \,(g^2+g^{\prime 2})^{1/2}}{
  g^2\,(1-2\, \delta^2) \csc 2\tilde\chi +(g^2+2 g^{\prime 2})\,\cot 2\tilde\chi}
% \nonumber\\[10pt]
\qquad
\text{with}\qquad\delta^2  = \frac{\q^2\,\gamma^2}{g^{2}}\coth^2 \frac{\langle\sigma\rangle}{M_p\sqrt{6}}.
\eea

\medskip\noindent
The masses of $Z$ boson ($m_Z$) and Weyl gauge field ($m_\w$) are then
found\footnote{If there is no mixing, $\chi=0$, then in eq.(\ref{ze}), also (\ref{U}),  $\zeta=0$,
  and  with $M_p$   of (\ref{planck}) and $h$ of (\ref{hw}) then
\bea
m_\w^2=\frac{3\q^2}{2}\gamma^2 M_p^2\,\Big[1+\sinh^2\!\frac{\langle\sigma\rangle}{M_p\sqrt{6}}\Big]
=\frac{\q^2}{4} \gamma^2 \Big[(1+\xi_h) \langle h\rangle^2  + \frac{\langle\phi\rangle^2}{\xi^2}\Big],
\quad
  m_Z^2\!=\! \frac32 (g^2\!+\! g^{\prime 2})  M_p^2 \sinh^2\!\frac{\langle\sigma\rangle}{M_p\sqrt 6}.
  \eea}
\medskip
\bea
m_{Z,\,\w}^2&=&\frac{3}{4} M_p^2 \sinh^2\frac{\langle\sigma\rangle^2}{M_p\sqrt 6}\,
\Big\{ g^2+ \frac12\sec^2\tilde\chi  \Big[2 \,g^{\prime 2} + 2\,\q^2 \gamma^2\,\coth^2
\frac{\langle\sigma\rangle}{M_p}\pm \sqrt{\cP}\Big]\Big\},
\\[4pt]
\text{where}&&
\cP=4\,g^{\prime 2} (g^2+g^{\prime 2})\,\sin^2 2\tilde\chi
+ \Big[ g^2\,(1- 2\, \delta^2)
+ (g^2+2 g^{\prime 2})\,\cos 2\tilde\chi\Big]^2.
\eea

\medskip\noindent
Since $\langle\sigma\rangle\ll M_p$ (see conditions after eq.(\ref{min}))
\medskip
\bea
m_Z^2=\frac14\, (g^2 +g^{\prime 2})\,\langle\sigma\rangle^2
\Big\{
1+\frac{\langle\si\rangle^2}{18 M_p^2}\Big[ 1-\frac{3 \,g^{\prime 2}}{\q^2} \sin^2\chi\Big]
+
\cO(\langle\sigma\rangle^4/M_p^4)
\Big\}.
\eea

\medskip
The factor in front is the mass of $Z$ boson (hereafter $m_{Z^0}$) in the SM;
$m_Z$ has a negligible correction from  Einstein gravity ($\propto\langle\sigma^2\rangle/M^2$).
But there is also a correction ($\propto\sin^2\chi/\q^2$) from  the Weyl field
i.e. due to deviations from Einstein gravity induced by Weyl geometry.
This can be significant and it {\it reduces}  $m_Z$ by a relative amount:
\medskip
\bea\label{eps}
\varepsilon\equiv
 \frac{\Delta m_Z}{m_{Z^0}}
=
-\frac{g^{\prime\, 2}\,\langle\sigma\rangle^2}{12\,M_p^2} \,\frac{\sin^2\chi}{\q^2}+
\cO\Big(\frac{\langle\sigma\rangle^4}{M_p^4}\Big)
=-\frac18 \Big(\frac{\langle\sigma\rangle}{m_\w}\Big)^2\,
(g^{\prime}\,\tan\tilde\chi)^2
+\cO\Big(\frac{\langle\sigma\rangle^4}{m_w^4}\Big).
\eea
%
%\medskip\noindent
In  the second step we replaced the mass of $\w$
and the definition of $\tilde\chi$ in eq.(\ref{bw}).

The effect in (\ref{eps}) is significant if  $\sin\chi/\alpha\gg 1$.  From the mass of $Z$ boson and
 with $\Delta m_Z$ at 1 $\sigma$ deviation, one has $\vert\varepsilon\vert\leq 2.3 \times 10^{-5}$, 
 then eq.(\ref{eps}) gives  a lower bound on the Weyl gauge coupling $\q$, for a given non-zero
 gauge kinetic mixing:
\bea\label{alphasin}
\q\geq 2.17\times 10^{-15} \sin\chi.
\eea
Note that for an arbitrary charge $d$ of the metric, the
results depending on $\q$ are modified by replacing $\q\ra  d\times  \q$.
In terms of the mass of $\w_\mu$ one finds
\bea\label{t}
\frac{m_\w}{\text{TeV}}\geq 6.35\times \tan\tilde\chi.
\eea
%\medskip
This gives a lower bound on the mass of the Weyl field in terms of the mixing
angle $\chi$ and $\gamma$. A larger $m_\w$ allows  a larger amount of mixing.
For  a mixing angle of e.g. $\tilde\chi=\pi/4$ then  $m_w\geq 6.35$ TeV.
Note that if there is no term  $(1/\eta)\,\tilde C_{\mu\nu\rho\sigma}^2$
in the original gravity action, then $\gamma=1$ and then  $\chi=\tilde\chi$.
Alternatively,  using the current lower  bound on the non-metricity scale
(represented by $m_\w$)  which is of the order of the TeV scale \cite{Latorre}, then
\bea
\tan\tilde\chi\leq 0.16
\eea
This is consistent with the non-metricity constraint.

These  bounds  are significant and affect other phenomenological studies.
To give an example, consider the impact of  $\w_\mu$  on the $g-2$ muon magnetic moment,
due to  the new coupling of $\w_\mu$  in  $\cL_f$, eq.(\ref{Lff}).
Using \cite{CB,Marciano} an estimate of the correction of  $\w_\mu$ to  $\Delta a_\mu$ is 
\bea
\Delta a_\mu\sim \frac{1}{12\pi^2}\, \frac{m_\mu^2}{m_\w^2} \, (g^\prime \tan\tilde\chi)^2
=2.56\times 10^{-13},
\eea
where we used constraints (\ref{eps}), (\ref{t}).
These  do not allow $\Delta a_\mu$  to account for the SM discrepancy with the experiment
\cite{Abi}; however, this discrepancy  may be only apparent, according to
lattice-based results \cite{szabo}.
One can also use these constraints  when studying  the role of
$\w_\mu$ for  phenomenology in other examples, such as the dark matter problem \cite{Tang},
in which case it may even provide a solution (of geometric origin!) to this problem;
other implications can be for example  in the  birefringence of the vacuum induced by  $\w_\mu$.
This can impact  on  the propagation of the observed  polarization of the gamma-ray
bursts \cite{Harko} or of the CMB \cite{Minami}.

\subsection{Inflation}\label{s2.8}

The  SMW model can have successful inflaton. For large $\sigma\sim M_p$, the Higgs
potential in (\ref{Vinflation})  can drive inflation \cite{Ross,Winflation,Winflation2}.
But who ``ordered'' the Higgs in the early Universe?  the Higgs could
initially  be produced  by the Weyl gauge boson fusion via its
coupling $\w_\mu \w^\mu H H^\dagger$ dictated by the symmetry, eq.(\ref{higgsR}).
This means, rather interestingly,  that the Higgs can be regarded  as having a {\it geometric
  origin}, just like  $\w_\mu$ which is part of the Weyl connection\footnote{
  In a sense this is also true for fermions, by subsequent Higgs decay
  (\ref{Y}), or for $B_\mu$  by Higgs-$\w_\mu\!\ra\! B_\mu$-Higgs.}.

As seen from (\ref{LL}) this coupling becomes $\w_\mu \w^\mu f(\sigma)$
with $\sigma$ the neutral Higgs.  But in   a Friedmann-Robertson-Walker
universe considered below,
 $g_{\mu\nu}\!=\!(1,-a(t)^2, -a(t)^2, -a(t)^2)$,  the vector field background
compatible with the metric is $\w_\mu(t)=0$ \cite{Winflation2}. The fluctuations of $\sigma$
and of (longitudinal component of) $\w_\mu$ do not mix since $\w_\mu(t)\delta\w^\mu\delta\sigma$ is
then vanishing. As a result, the  single-field inflation
formalism in the  Einstein gravity applies, with $\sigma$ as the inflaton.
Since $M_p$ is simply the scale of Weyl gauge symmetry breaking, $\sigma>M_p$ is natural.

The predictions of the Higgs inflation are then \cite{Winflation,Winflation2}
%\medskip
\be\label{rn}
r=3 \,(1-n_s)^2- \frac{16}{3}\, \xi_h^2 +\cO(\xi_h^3).
\ee
%
%\medskip\noindent
Here $r$ is the tensor-to-scalar ratio and $n_s$ is the scalar spectral index.
Up to small corrections from $\xi_h$ that can be neglected for $\xi_h<10^{-3}$,
the above dependence $r=r(n_s)$ is similar to that in the
Starobinsky model \cite{Sta} of inflation where $r=3 (1-n_s)^2$. 
For mildly larger $\xi_h \sim 10^{-3} - 10^{-2}$, eq.(\ref{rn})
departs from the Starobinsky model prediction
and $r$ is mildly reduced relative to its value in the
Starobinsky case, for  given  $n_s$.
These results  require  a hierarchy $\lambda\ll \xi_h^2\, \xi^2$  which may be
respected by a sufficiently small $\lambda$ and\footnote{From
  the normalization of the CMB anisotropy one also finds that  
  $\xi^2< 1.45 \times 10^{-9}$ \cite{Winflation}.}
  $\xi_h\sim 10^{-3} - 10^{-2}$.

A relatively very small  $\lambda$ means that it is actually
the squared term in (\ref{Vinflation}) that is multiplied by $\xi^2$ (see also (\ref{neg}))
that is mostly  responsible for inflation; this is  the Stueckelberg field
contribution, ultimately due to the initial term
$\phi^4$ arising  from the initial {\it quadratic
curvature $(1/\xi^2)\,\tilde R^2$ term} in action (\ref{higgsR}).
This  then explains the close similarities to  the Starobinsky $R^2$-inflation.
Thus, we actually have  a Starobinsky-Higgs  inflation. The initial Higgs field $h$
(which has  $\xi_h\!\not=\!0$)  does play a role as it brings  a minimum in\footnote{
  The Higgs and  Starobinsky/$R^2$ inflation usually mix
 at a quantum level \cite{Twoloop}.} $\hat V(\sigma)$ of (\ref{Vinflation}).
In conclusion,  a negligible $\lambda$ is required for successful inflation
(as the numerical values of $r$ below also show it).
This is consistent with SM prediction for $\lambda$ at the high scales, while 
a value of $\lambda$ at the EW scale as  in the SM
can then be induced  by the SM quantum corrections.

The  numerical results give that for $N=60$ efolds and with
$n_s=0.9670\pm 0.0037$ at $68\%$ CL (TT, TE, EE+low E + lensing + BK14 + BAO) \cite{planck2018}
then \cite{Ross,Winflation,Winflation2}\footnote{Our results quoted above from \cite{Winflation,Winflation2}
  were obtained from
  potential $\hat V(\sigma)$ of eq.(\ref{Vinflation}) and they agree both analytically and numerically
  to those in \cite{Ross} obtained by a different method using a two-field analysis.}
%\medskip
\be
0.00257\leq r\leq 0.00303, \quad (n_s\,\,\,\textrm{at}\,\,\,68\% \,\,\textrm{CL})
\ee
\be
0.00227\leq r\leq 0.00303, \quad (n_s\,\,\,\textrm{at}\,\,\,95\% \,\,\textrm{CL})
\ee
%\medskip\noindent
The case of Starobinsky model for $N=60$ corresponds to the upper limit of $r$ above
and is reached for the smallest $\xi_h$, when this limit is saturated, according to
relation (\ref{rn}).

The small value of $r$ found above may be reached by the next
generation of CMB experiments CMB-S4 \cite{CMB1,CMB2}, LiteBIRD
\cite{litebird,CMB3}, PICO \cite{CMB4},
PIXIE \cite{Pixie} that have sensitivity to $r$ values as low as $0.0005$.
Such sensitivity will be able to test this  inflation model
and to distinguish it from  other models.
For example,  similarly small but distinct values of $r$ are
found in other models with Weyl gauge symmetry
\cite{Ghilen2,Winflation2} based on the Palatini approach to 
gravity action (\ref{inA}) used in this paper; however
these models do not respect relation (\ref{rn}) and the slope of the curve $r(n_s)$ is
different, due to their different vectorial non-metricity.
The above experiments also have the sensitivity to distinguish inflation in this model from
the Starobinsky model for  $\xi_h\sim 10^{-2}$ when the curve $r(n_s)$ is shifted by $\xi_h$
below that of the Starobinsky model, towards smaller $r$ (for fixed $n_s$).

\section{SMW and its properties}\label{s3}

\vspace{-0.1cm}
In this section we discuss some features of our model
and  the differences from other SM-like models with local scale invariance.
The main aspect of our  model is that scale symmetry is {\it gauged},  eq.(\ref{WGS}).
The Weyl gauge symmetry is  not only a symmetry of the action  but also of
the underlying Weyl  geometry;
indeed, the Weyl connection $\tilde \Gamma$ (\ref{AGamma}) is invariant under (\ref{WGS})
and the same is true about the
Weyl spin connection (\ref{WSC}).
This adds consistency to SMW  and distinguishes  it from models with
an action that is Weyl or conformal invariant (with no  $\w_\mu$),
built in a (pseudo-)Riemannian space; their Levi-Civita  connection and thus their
underlying  geometry do not have the  symmetry of the action - which may be a concern.

An important feature of the SMW  is the  spontaneous breaking
of Weyl gauge symmetry even  in the {\it absence of matter}, as seen  in 
section~\ref{s2.1}.  Hence, this breaking is
 of geometric origin.  This is different from previous models
with this symmetry \cite{Dirac,Kugo,Smolin,Cheng,Fulton,Wheeler,Moffat1,
  Nishino,Ohanian,Moffat2,Tann,Guendelman,ghilen,Quiros1,Quiros2,pp1,pp2,pp3,pp4,pp5} where
some scalar fields were  introduced  ``ad-hoc''  to induce spontaneous breaking
of their symmetry and to generate $M_p$ and Einstein action by a $\phi^2 R$ term.
In the SMW  the necessary  scalar field ($\phi$) is ``extracted'' from the geometric
$R^2$-term, plays the role of the Stueckelberg field and is eaten by $\w_\mu$
which becomes massive.  This was possible since the model was quadratic
in curvature - this is another difference from models
\cite{Dirac,Kugo,Smolin,Cheng,Fulton,Wheeler,Moffat1,Nishino,Ohanian,Moffat2,Tann,Guendelman,
  ghilen,Quiros1,Quiros2,pp1,pp2,pp3,pp4,pp5} which are linear in $R$.
Therefore, the Einstein-Proca action and the Planck scale  emerge in  the broken phase
of the SMW.

The breaking of the  Weyl gauge symmetry is accompanied
by  a change of the underlying geometry.
When massive  $\w_\mu$ decouples at some (high) scale,  the Weyl connection becomes Levi-Civita,
so Weyl geometry becomes Riemannian and  the theory 
is then metric\footnote{A similar Weyl gauge
  symmetry breaking and change of  geometry exists in
a Palatini version  \cite{Ghilen2,Winflation2}.}.
Thus, the breaking of the symmetry in Section~\ref{s2.1}
(see \cite{Ghilen1}) is not just a  result of a
 ``gauge fixing'' to the Einstein frame,
as it happens in Weyl or conformal theories with no $\w_\mu$; it
 is  accompanied by the  Stueckelberg
 mechanism   and by a change of the underlying geometry\footnote{
  An aspect of models with  Weyl gauge symmetry  
  relates to  their  geodesic completeness, see \cite{Ohanian,Quiros1}.
  In conformal/Weyl invariant models (without $\w_\mu$) this aspect
  seems possible  in  the (metric)  Riemannian spacetime   where geodesic
  completeness or incompleteness  is related to a gauge choice
    (and singularities due to an unphysical conformal frame)
  \cite{narlikar,modestoJCAP,modesto1605}.
In models in Weyl geometry,
the geodesics  are determined by the affine structure. 
 Differential geometry demands  the existence of the Weyl gauge field \cite{Ehlers}
  for the construction of the affine connection, % 
  because this ensures that geodesics are  invariant (as necessary on  physical grounds,
  the parallel transport of a vector should not depend on the  gauge choice).
  Hence the Weyl gauge field/symmetry may  actually be  required! % 
  After the  breaking of this symmetry, $w_\mu$  decouples, we return to
  (pseudo)Riemannian geometry and geodesics  are then given by extremal proper time condition.
   Since a dynamical $\w_\mu$ also  brings in non-metricity, geodesic completeness
   seems related to non-metricity.}.

 The SMW   avoids some situations  present in 
 interesting models  with  local scale invariance  (without $\w_\mu$),
 like a negative kinetic term of the scalar field \cite{Bars0} (also \cite{Bars1,Bars2,Kallosh-2}),
 or an imaginary vev  \cite{tH0,tH3} of the scalar that generates\footnote{It seems to us
   this means a negative $\Sigma$ and therefore a metric signature change in transformation (\ref{WGS}).}
 $M_p$. Such situations may be  a cause of concern according to
  \cite{Ohanian,Quiros1}.
 Gauging the scale symmetry  avoids such situations - in SMW this  scalar field
plays the role of  a would-be Goldstone of the Weyl gauge symmetry 
(eaten by $\w_\mu$).  See also eq.(\ref{ST}) where
the (negative) kinetic term in the first square bracket is cancelled by that in the
second square bracket corresponding to a Stueckelberg
mechanism\footnote{This  Stueckelberg mechanism  may  apply to more general
  metric affine theories studied in \cite{P3}.}.

In  local scale invariant models (without $\w_\mu$) 
the associated current can be  trivial, leading to so-called ``fake conformal symmetry''
\cite{J1,J2};  in the SMW the current is non-trivial even in the absence of matter \cite{Ghilen1}
due to  dynamical  $\w_\mu$.
If  $\w_\mu$  were not dynamical ($F_{\mu\nu}=0$) it  could be integrated out algebraically 
to leave a  local scale-invariant action \cite{Ghilen1,Ghilencea:2022lcl};
in this case   Weyl geometry would be integrable and metric, see e.g.
\cite{Quiros1,Quiros2}. But since $\w_\mu$  is dynamical,
the theory  is also non-metric. This non-metricity
would indeed be a physical problem if $\w_\mu$ were massless
(assuming this, non-metricity of a theory
was used as an argument against such theory by Einstein\footnote{Actually, a similar situation
  exists \cite{Winflation2,Ghilen2}  in quadratic  gravity in  Palatini approach  due to
  Einstein \cite{E2}. }
\cite{Weyl1}).
However, non-metricity became here an advantage,  since  Weyl geometry
with dynamical $\w_\mu$ enabled the Stueckelberg breaking mechanism,
$\w_\mu$ acquired a mass (above current non-metricity bounds
\cite{Latorre}),  and the  Einstein-Proca action was naturally obtained in the broken phase.

The SMW   differs from the SM with conformal symmetry of \cite{Nicolai} or
\cite{tH0, tH3} and from conformal gravity models \cite{Mannheim,Faria1,Faria2} formulated in the
(pseudo)Riemannian space and based on $C_{\mu\nu\rho\sigma}^2$ term; these models
have  metric geometry and do not have a gauged scale symmetry;
in our case the $C_{\mu\nu\rho\sigma}^2$ term is largely spectator
and may be absent in a first instance; it was included because its Weyl geometry
counterpart gave a correction to $\alpha$ and  it is needed at a quantum level. 
And unlike the  conformal gravity action \cite{Kaku} which {\it is metric}, 
the SMW  has a gauge kinetic term for  the Weyl field which 1) makes the geometry non-metric and
2) breaks the special
conformal symmetry; this symmetry and non-metricity do not seem compatible.

Concerning  the quantum calculations in the SMW, 
 one could try to use the ``traditional'' dimensional regularization (DR),
but that breaks explicitly  the Weyl gauge symmetry by the presence of the
subtraction scale ($\mu$).  One should 
use instead a regularisation similar to \cite{Englert} that preserves  Weyl gauge symmetry
at the quantum level.  This is possible by using our Stueckelberg
field $\phi$ as a {\it field-dependent regulator},   to replace the subtraction scale $\mu$
generated later by $\mu\sim \langle\phi\rangle$ (after symmetry breaking). This would allow the
computation of the quantum corrections without explicitly breaking the Weyl gauge symmetry\footnote{
  A similar approach exists  in the global case \cite{MS,MS2,quantumG2,quantumG4}.}.

 It is interesting to study the renormalizability of  the Weyl quadratic gravity and of the SMW.
 The usual (metric) quadratic gravity theory in the (pseudo-)Riemannian case is
 known to be renormalizable but not unitary due to the massive spin-2 ghost  \cite{Stelle}.
Considering now the Weyl quadratic gravity alone, note that for computing the quantum corrections
eq.(\ref{EP}) is not appropriate since this is the (non-renormalizable)
{\it unitary gauge} of Weyl gauge symmetry. Therefore, one should consider
  computing the necessary quantum corrections in the symmetric phase, for example in $\cL_0$ of  eq.(\ref{alt}).
Note that no higher order  operators are allowed by the symmetry in eq.(\ref{inA}), (\ref{alt}),
since there is no initial mass scale to suppress them, and this is an argument in favour of
its renormalizability.
Finally, regarding the SMW itself,
in a Riemannian  notation it simply has  an additional (anomaly-free)
 Weyl gauge field which becomes massive by the Stueckelberg mechanism which
cannot affect renormalizability; 
naively,  one then  expects the SMW  be renormalizable.

 \section{Conclusions}\label{s4}

Since the SM with a vanishing Higgs mass parameter is scale invariant,
it is natural to study the effect of  this symmetry. This is relevant 
for physics at high scales or in the early Universe, where this symmetry seems natural.
Since a {\it global} scale symmetry does not survive  black-hole physics, 
we  explored the possibility that the SM has a {\it gauged} scale symmetry.
The natural framework is the  Weyl geometry where
this symmetry is {\it built in}. Hence, we considered the SM in Weyl  geometry.
This embedding is both natural and minimal i.e.  {\it no new degrees of freedom} were needed or added
beyond those of the SM and of Weyl geometry.

The model has the  special feature that both  the action
{\it and} its underlying geometry
(connection $\tilde \Gamma$  and spin connection $\tilde w_\mu^{ab}$)
are  Weyl gauge invariant.  This adds consistency to the model and
distinguishes it from previous  SM-like models with local scale symmetry, built
in a (pseudo-)Riemannian geometry whose connection is not
local scale invariant.

The SMW  model has another  attractive feature.
In Weyl  geometry there exists a (geometric) Stueckelberg mechanism in which
this symmetry is spontaneously broken. The
Weyl quadratic  gravity associated to this geometry  is
broken spontaneously to the Einstein-Proca action of $\w_\mu$.
The Stueckelberg field $\phi$ has a geometric origin, being ``extracted'' from $\tilde R^2$ 
in the Weyl action,  and is subsequently eaten by $\w_\mu$.
Once the  Weyl gauge field decouples, the
Weyl connection becomes Levi-Civita and Einstein gravity is recovered.
The Planck scale and a positive cosmological constant are both generated by
the Stueckelberg field vev. 
Also,  the  mass term of the Weyl field is on the Weyl geometry side
just   a Weyl-covariant kinetic term of the same Stueckelberg field.
These aspects  relate symmetry breaking and thus
mass  generation to a geometry change (from Weyl to Riemannian)
 which is itself related to the  non-metricity induced by dynamical $\w_\mu$.

The  SMW gauge group is a direct product of the SM gauge group and $D(1)$ of
the Weyl gauge symmetry, both broken spontaneously.
Usually, of the SM spectrum only the Higgs field ($\sigma$) has a coupling to $\w_\mu$,
the term $\q^2\w_\mu \w^\mu \sigma^2$.
The   Weyl gauge symmetry 
can protect the Higgs mass at a quantum level,  if
this symmetry is broken at a low scale. The breaking scale is set by $m_\w\sim \q M_p$,
and if the Weyl gauge coupling $\q\!\ll\! 1$, then $m_\w$ can be light, few TeV 
(which is the current lower bound on non-metricity).
The mass of $\w_\mu$  is then the highest physical
scale for the low-energy observer and  quantum corrections
to $m_\sigma^2$ will appear as $\propto m_\w^2$.
Above $m_\w$ the  gauged scale  symmetry is restored, 
together with its UV protection  (for the Higgs) not affected by the
spontaneous breaking. Hence Weyl gravity and its gauged
scale symmetry  could give
a solution to the hierarchy problem.
And since $m_\w$ also sets the non-metricity scale, 
then the  hierarchy problem and non-metricity scale may be related!

The fermions  can  acquire a   coupling $(Y g^\prime\tan\tilde\chi)$
to $\w_\mu$ only in the case of a small
kinetic mixing ($\tilde\chi$) of the gauge fields of $U(1)_Y\times D(1)$,
if this mixing is not forbidden by a discrete symmetry.
Due to such mixing part of Z boson mass is not due to the Higgs mechanism,
but  to the mixing of $Z$ with the massive Weyl  field
which has a Stueckelberg mass; hence, part of $Z$ mass has a geometric origin,
due to a departure from  the pseudo-Riemannian geometry and Einstein gravity.
Since  $m_Z$ is accurately measured,  one finds 
bounds on the Weyl gauge coupling and the mass of $\w_\mu$, for a given amount of
kinetic mixing.
We showed how these bounds can be  used in other  phenomenological studies.
If $\w_\mu$ is light (few TeV, $\q\ll 1$) its effects may be amenable to experimental
tests, with  consequences for phenomenology
e.g.: $\w_\mu$  as a dark matter candidate, the vacuum birefringence, etc.

The SMW has  successful inflation.
Intriguingly, in the early Universe the  Higgs may be produced via Weyl vector fusion,
thus having a geometric origin.
With $M_p$  a simple phase transition scale in Weyl gravity, Higgs field values larger than
$M_p$  are natural. Note that while  the inflationary potential is that of the  Higgs, due
to its mixing with $\phi$, it is ultimately a  contribution to this
potential from the initial scalar mode ($\phi$) in the $\tilde R^2$  term that is actually
responsible for
inflation. This explains the  close similarities to the Starobinsky $R^2$-inflation.
 With the scalar spectral index  $n_s$ fixed to its measured value,  the
tensor-to-scalar ratio $0.00227\leq r\leq 0.00303$. 
Compared to the Starobinsky model, the curve $r(n_s)$  is similar but shifted to smaller $r$
(for same $n_s$)  by the  Higgs  non-minimal coupling ($\xi_h$) to   Weyl geometry.
These results of the SMW model  deserve further investigation.

%\newpage
 \section*{Appendix}

\def\theequation{A-\arabic{equation}}
\def\thesubsection{A}
\setcounter{equation}{0}
\def\thefigure{A-\arabic{figure}}
\def\thelabel{A}

\subsection{Brief guide to  Weyl conformal geometry}
\label{AA}

Weyl conformal geometry is defined by equivalent classes of ($g_{\mu\nu}, \w_\mu$) of the metric
and Weyl gauge field ($\w_\mu$) related by Weyl gauge transformations:
\medskip
\bea\label{WGS2}
\!\!\!\hat g_{\mu\nu}\!=\!\Sigma^d %\Omega^2
\,g_{\mu\nu},\quad
\sqrt{\hat{g}}=\!\Sigma^{2 d} \sqrt{g},
\quad\,
\hat\w_\mu\!=\!\w_\mu -\frac{1}{\q}\, \partial_\mu\ln\Sigma, %\Omega^2
\quad\,
\hat e_\mu^{\,\,a}=\!\Sigma^{d/2} 
\, e_\mu^{\,\,a},
\quad\,
\hat e_a^\mu=\!\Sigma^{-d/2}\,e_a^\mu
\eea

\medskip\noindent
where $d$ is the Weyl weight (charge) of  $g_{\mu\nu}$ and $\alpha$ is the Weyl gauge coupling.
Various conventions exist in the literature for  $d$ e.g. $d=1$ in \cite{Smolin}
and $d=2$ in \cite{Kugo}. 
The latter  may be more motivated since from the relation
$d s^2=g_{\mu\nu} dx^\mu dx^\nu$ with $d x^\mu$ and $d x^\nu$ fixed under (\ref{WGS2}) the
metric $g_{\mu\nu}$  transforms like $ds^2$. In the text we used $d=1$,
but our results can be immediately
changed to arbitrary $d$ by simply rescaling the coupling in our results
  $\q \ra \q\,\times d$.

 The Weyl gauge field is  related to the Weyl
 connection ($\tilde\Gamma$) which  is the solution of
 \bea\label{nablag}
\tilde\nabla_\lambda  g_{\mu\nu}=- d\, \q\, \w_\lambda g_{\mu\nu}
\eea
where $\tilde\nabla_\mu$ is defined by  $\tilde\Gamma_{\mu\nu}^\lambda$
\bea\label{eeee}
\tilde\nabla_\lambda g_{\mu\nu}=\partial_\lambda g_{\mu\nu}- \tilde \Gamma^\rho_{\mu\lambda}
g_{\rho\nu} -\tilde\Gamma^\rho_{\nu\lambda}\,g_{\rho\mu}.
\eea

\medskip\noindent
Eq.(\ref{nablag}) says that Weyl geometry is {\it non-metric}; it may be written as $(\tilde \nabla_\lambda+d\,\q\,\w_\lambda)\,g_{\mu\nu}=0$
as in a metric case, indicating that one can use metric formulae in which replaces
the partial derivative $\partial_\lambda$  acting on a field, metric, etc, by a Weyl-covariant counterpart
as in:
\bea\label{cov}
\partial_\lambda \ra \partial_\lambda+ \rm{weight}\times \q\times \w_\lambda,
\eea
where 'weight' is  the corresponding Weyl charge (of the  field, metric, etc).
We  use this later.

The solution $\tilde \Gamma$ to (\ref{nablag}) is found using cyclic
permutations of the indices and combining the equations so
obtained, or by simply using (\ref{cov}) in Levi-Civita connection ($\Gamma$),
to find
\bea\label{AGamma}
\tilde \Gamma_{\mu\nu}^\lambda=
\Gamma_{\mu\nu}^\lambda+\q \,\frac{d}{2} \,\Big[\delta_\mu^\lambda\,\w_\nu
+\delta_\nu^\lambda\, \w_\mu - g_{\mu\nu} \,\w^\lambda
\Big],
\eea

\medskip\noindent
where $\Gamma_{\mu\nu}^\lambda$ is the usual Levi-Civita connection
\bea
\Gamma_{\mu\nu}^\lambda=\frac12\,g^{\lambda\rho} (\partial_\mu g_{\rho\nu}+
\partial_\nu g_{\rho\mu}-\partial_{\rho} g_{\mu\nu}).
\eea

\medskip\noindent
$\tilde \Gamma$  is invariant under (\ref{WGS2}) as one can easily check.
Conversely, one may actually derive
the transformation of the Weyl gauge field in (\ref{WGS2}) by imposing that $\tilde \Gamma$
be invariant under the metric change in (\ref{WGS2}), since parallel
transport should be independent of the  gauge choice.
Taking the trace  in the last equation and denoting $\Gamma_\mu\equiv \Gamma_{\mu\lambda}^\lambda$ and
 $\tilde\Gamma_\mu\equiv \tilde\Gamma_{\mu\lambda}^\lambda$ then
\bea
\tilde \Gamma_\mu=\Gamma_\mu + 2 d \,\q\,\w_\mu.
\eea
Thus, the Weyl gauge field can be thought of as the trace of the  departure
of  the Weyl connection from the Levi-Civita connection.
Using $\tilde\Gamma$ one computes the scalar and tensor curvatures of Weyl geometry,
using formulae similar to those in Riemannian case but with $\tilde\Gamma$ instead of $\Gamma$. For example
%\medskip
\bea
\tilde R^\lambda_{\mu\nu\sigma}=
\partial_\nu \tilde\Gamma^\lambda_{\mu\sigma}
-\partial_\sigma \tilde\Gamma^\lambda_{\mu\nu}
+\tilde\Gamma^\lambda_{\nu\rho}\,\tilde\Gamma_{\mu\sigma}^\rho
-\tilde\Gamma_{\sigma\rho}^\lambda\,\tilde\Gamma_{\mu\nu}^\rho,
\qquad
\tilde R_{\mu\nu}=\tilde R^\lambda_{\mu\lambda\sigma},
\qquad
\tilde R=g^{\mu\sigma}\,\tilde R_{\mu\sigma}.
\eea

\medskip\noindent
After some algebra one finds
\medskip
\bea\label{tRmunu}
\tilde R_{\mu\nu}&=&R_{\mu\nu}
+\frac 12\, (\q d) (\nabla_\mu \w_\nu-3\,\nabla_\nu \w_\mu - g_{\mu\nu}\,\nabla_\lambda \w^\lambda)
+\frac12 \,(\q d)^2\, (\w_\mu \w_\nu -g_{\mu\nu}\,\w_\lambda \w^\lambda),\qquad
\\[8pt]
&& \qquad\qquad\qquad\qquad
\tilde R_{\mu\nu}-\tilde R_{\nu\mu}=2 \,d\q\, F_{\mu\nu},
\\[8pt]
\tilde R&=& R-3 \,d\,\q\,\nabla_\mu\w^\mu-(3/2)\,(d\,\q)^2 \,\w_\mu\,\w^\mu,
\label{tR}
\eea

\medskip\noindent
where the rhs is in a Riemannian notation, so $\nabla_\mu$ is given by
the Levi-Civita connection ($\Gamma$).

An important property is that $\tilde R$  transforms covariantly under (\ref{WGS2})
\bea
\hat{\tilde R}=(1/\Sigma^d)\,\tilde R,
\eea
which follows from the transformation of $g^{\mu\sigma}$ that enters its
definition above and from the fact that $\tilde R_{\mu\nu}$ is invariant (since
$\tilde \Gamma$ is so). Then  the term $\sqrt{g}\, \tilde R^2$
is Weyl gauge invariant.

In Weyl geometry one can also define a Weyl tensor $\tilde C_{\mu\nu\rho\sigma}$
that is related to that in Riemannian geometry $C_{\mu\nu\rho\sigma}$ as follows
\bea\label{tC2}
\tilde C_{\mu\nu\rho\sigma}=
C_{\mu\nu\rho\sigma}
- \frac{\q\,d}{4} \,(g_{\mu\rho} F_{\nu\sigma}
+g_{\nu\sigma} F_{\mu\rho}
-g_{\mu\sigma} F_{\nu\rho}
-g_{\nu\rho} F_{\mu\sigma})
+\frac{\q\,d}{2} F_{\mu\nu} g_{\rho\sigma}
\eea
which gives \cite{Tann}
\bea\label{tC}
\tilde C_{\mu\nu\rho\sigma}^2=C_{\mu\nu\rho\sigma}^2+\frac{3}{2}\, (\q\,d)^2 \, F_{\mu\nu}^2,
\eea

\medskip\noindent
used in the text, eq.(\ref{inA}).
$\sqrt{g}\,\tilde C_{\mu\nu\rho\sigma}^2$ and its above separation are invariant under (\ref{WGS2}).

To introduce the  Weyl spin connection,  consider first the spin connection in the
Riemannian geometry 
\bea\label{spin}
s_\mu^{\,\,ab}=\frac12\,\Big[
e^{\nu a} (\partial_\mu e_\nu^b -\partial_\nu e_\mu^b)
-
e^{\nu b}\,(\partial_\mu e_\nu^a -\partial_\nu e_\mu^a)
-
e^{\rho a}\, e^{\sigma b}\, e_\mu^c (\partial_\rho e_{\sigma c} -\partial_\sigma e_{\rho c})
\Big].
\eea

\medskip\noindent
One verifies that an equivalent form is
\bea
s_\mu^{\,\,ab}=-e^{\lambda b}(\partial_\mu e_\lambda^a - \Gamma_{\mu\lambda}^\nu \,e_\nu^a).
\eea
Under a transformation of the metric (\ref{WGS2})
\bea
\hat s_\mu^{\,\,ab}=s_\mu^{\,\,ab}+(e_\mu^a \,e^{\nu b}-e_\mu^b\,e^{\nu a})\,
\partial_\mu\ln\Sigma^{d/2}.
\eea

\medskip
For the Weyl geometry spin connection, one simply replaces the partial derivative in eq.(\ref{spin}) 
by a Weyl-covariant derivative that takes into account the charge of the field on
which it acts (\ref{cov}).
For the spin connection   $\partial_\mu e_\nu^b \ra [\partial_\mu +(d/2) \q\,\w_\mu] \,e_\nu^b$
since according to (\ref{WGS2}) $e_\nu^b$ has Weyl weight $d/2$. Using this replacement
in (\ref{spin}) we find the spin connection $\tilde s_\mu^{\,\,ab}$ in Weyl geometry
\bea\label{WSC}
\tilde s_\mu^{\,\,ab}=s_\mu^{\,\,ab}+(d/2)\,\q\,(e_\mu^a\,e^{\nu b}-e_\mu^b\,e^{\nu a})\,\,\w_\nu.
\eea
%\medskip\noindent
Under  transformation (\ref{WGS2}) one checks that $\tilde s_\mu^{ab}$ is 
invariant, similar to Weyl connection $\tilde \Gamma$.

Let us now consider  matter fields and find their  charges in Weyl geometry
by demanding  that:  a) their Weyl-covariant derivatives transform
under (\ref{WGS2}) like the fields themselves and b) that their kinetic terms  be
invariant.
More explicitly, take the kinetic term for a scalar of charge $d_\phi$:
$\sqrt{g} (\tilde D_\mu \phi)^2$ where $\tilde D_\mu$ is the Weyl-covariant
derivative which we demand it transform under (\ref{WGS2}) just like the
scalar field itself, i.e. it has same charge $d_\phi$. From the invariance
of this action under (\ref{WGS2}) one has that $d_\phi=-d/2$. 
The Weyl covariant derivative is then found according to (\ref{cov})
and  the kinetic term is
%\medskip
\bea
L_\phi=\sqrt{g} \, g^{\mu\nu} \tilde D_\mu\phi \tilde D_\nu \phi, \qquad
\tilde D_\mu\phi=(\partial_\mu -d/2 \, \q\, \w_\mu)\,\phi.
\eea
%
%\medskip\noindent
with $L_\phi$ invariant,  while  $\phi$ transforms as
\bea
\quad\hat\phi ={\Sigma^{-d/2}}\,\,\phi.\qquad\quad
%\hat\psi ={\Sigma^{-3 d/4}}\,\,\psi,
\label{WGS2p}
\eea

For a fermion $\psi$ the Weyl charge is found in a similar way, by using (\ref{cov}) to write
their Weyl covariant derivative, hence the action has the form
\medskip
\bea\label{psi}
L_\psi=\frac{i}{2}\,\sqrt{g}\,\overline\psi\,\gamma^a \,e^\mu_a \,\tilde\nabla_\mu \psi+h.c.,
\qquad
\tilde\nabla_\mu\psi=\Big[
\partial_\mu +d_\psi\,\q\,\w_\mu +\frac12 \,\tilde s_\mu^{\,\,ab}\,\sigma_{ab}
\Big]\psi
\eea

\medskip\noindent
where $\sigma_{ab}=1/4[\gamma_a,\gamma_b]$.
Since we saw earlier that $\tilde s_\mu^{\,\,ab}$ is Weyl gauge invariant
then the above derivative $\tilde\nabla_\mu\psi$ transforms covariantly just like a fermion field itself of
charge $d_\psi$. From the structure of the kinetic term and its invariance it follows
that $d_\psi=-3 d/4$ so, under (\ref{WGS2})
\bea\label{WGS3}
\hat\psi=\Sigma^{-3 d/4}\,\psi.
\eea
With this charge and using  (\ref{psi}), (\ref{WSC}) one shows that $\w_\mu$ cancels out:
\medskip
\bea
\gamma^a\, e_a^\mu\,\tilde\nabla_\mu\psi=\gamma^a \,e_a^\mu\,
\Big[\partial_\mu+\frac12 \,s_\mu^{\,\,ab} \,\sigma_{ab}\Big]\,\psi.
\eea

\medskip\noindent
Hence,  the fermionic kinetic term has the same form as in the Riemannian geometry
\bea\label{psi2}
L_\psi=\frac{i}{2}\sqrt{g}\,\overline\psi\gamma^a \,e^\mu_a \,\nabla_\mu \psi+h.c.,
\qquad
\nabla_\mu\psi=\Big[\partial_\mu +\frac12\, \, s_\mu^{\,\,ab}\,\sigma_{ab}\Big]\psi,
\eea
used in Section~\ref{s2.3}.
Eqs.(\ref{WGS2}), (\ref{WGS2p}), (\ref{WGS3}) define the Weyl gauge transformation
in the presence of matter, as introduced in the text, eq.(\ref{WGS}).
For more information see also \cite{Kugo,Tann}.

\bigskip

\def\theequation{B-\arabic{equation}}
\def\thesubsection{B}
\setcounter{equation}{0}
\def\thefigure{B-\arabic{figure}}
\def\thelabel{B}

\subsection{Weyl quadratic gravity: equations of motion and gauge fixing}
\label{apb}

Here we present the equations of motion of  $\cL_0$ of eq.(\ref{alt}) and derive some results 
that were used in the text, section~\ref{s2.1}.
Variation  of $\cL_0$ with respect to $g^{\mu\nu}$ gives
\medskip
\bea\label{eqg}
\frac{1}{\sqrt{g}}
\frac{\delta \cL_0}{\delta g^{\mu\nu}}
&=&\frac{1}{12\,\xi^2 }\Big\{
-  \phi^2\Big( R_{\mu\nu}- \frac12\,g_{\mu\nu} \,R\Big)
- \Big(  g_{\mu\nu}\Box-\frac12 (\nabla_\mu \nabla_\nu+\nabla_\nu\nabla_\mu)\Big) \phi^2
\nonumber\\
&-&\frac{3\q^2}{4} \phi^2 \Big( g_{\mu\nu}\,\w^\rho\,\w_\rho - 2 \w_\mu\,\w_\nu\Big)
-\frac{3\,\q}{2} \Big( \w_\nu\nabla_\mu+\w_\mu\nabla_\nu - g_{\mu\nu}\,\w^\rho\nabla_\rho\Big)\phi^2 \Big\}
\nonumber\\[4pt]
&+&\frac12 \,g_{\mu\nu}\,V
+\frac12\,\Big(\frac14 \, g_{\mu\nu}\,F_{\alpha\beta}\,F^{\alpha\beta} -g^{\alpha\beta}\,F_{\mu\alpha}
F_{\nu\beta}\Big).
\eea
where we denoted $V\equiv \phi^4/(4!\,\xi^2)$.
Taking the trace of (\ref{eqg})
\bea\label{eqt}
\frac{1}{12}\, \frac{1}{\xi^2}\,
\Big[\phi^2\, R - 3 \,\Box\phi^2
-\frac32 \q^2\,\phi^2 \,\w_\rho\,\w^\rho
+3 \q\,\w^\rho\nabla_\rho  \,\phi^2\Big]
+2 V=0.
\eea

\medskip\noindent
The equation of motion of $\phi$ (multiplied by $\phi$) gives
\bea\label{eqf}
\frac{1}{12}\frac{1}{\xi^2}\, \phi^2 \Big[ R
-\frac32 \q^2 \,\w_\rho\,\w^\rho
-3 \q\,\,\nabla_\rho\,\w^\rho\Big] +
\frac12\,\phi \,\frac{\partial V}{\partial \phi}=0
\eea

\medskip\noindent
This is just another form of  $\phi^2=-\tilde R$, see (\ref{def}), that 
was used to linearise action (\ref{inA}) into (\ref{alt}).

Subtracting  (\ref{eqf}) from (\ref{eqt}) and using the definition of $V$ then
\bea\label{box}
-\frac{1}{4\xi^2}\, \nabla_\rho\,\big[\big(\nabla^\rho-\alpha\omega^\rho\big)\,\phi^2\big]=0.
\eea
Thus, there is a conserved current which can also be seen from the
equation of motion of $\w_\mu$:
\bea\label{eqw}
\frac{\q^2}{4 } \frac{\phi^2}{\xi^2}
\,\w^\rho -\frac{\q}{4\,\xi^2} \,g^{\rho\sigma} \,\nabla_\sigma \phi^2
+\nabla_\sigma F^{\sigma\rho}=0.
\qquad\qquad\quad
\eea
By applying $\partial_\rho$ to (\ref{eqw})
and using $\sqrt{g}\nabla_\sigma F^{\sigma\rho}\!=\!\partial_\sigma (\sqrt{g}\,F^{\sigma\rho})$ with $F^{\sigma\rho}$
antisymmetric then
% \medskip
\bea
\nabla_\mu J^\mu=0,\quad\textrm{where}\quad
J^\mu
=-\frac{\q}{4\,\xi^2} g^{\mu\nu}\big(\partial_\nu-\q\,\w_\nu\big) \phi^2.
\eea
This was used in eq.(\ref{jj}).
When $\phi$ acquires a vev (at quantum level, etc)
then $\nabla_\mu J^\mu\!\!=0$  becomes $\nabla_\mu \w^\mu\!=\!0$. This is the gauge fixing
 for the massive gauge field $\w_\mu$  in action (\ref{EP}).

Finally, after the decoupling of massive $\w_\mu$ from the Einstein-Proca $\cL_0$
of eq.(\ref{EP}) (together with (\ref{la})), the equation of motion for $g^{\mu\nu}$ gives,
after taking the trace
\bea\label{ff}
R=-4 \Lambda.
\eea
This equation is also  seen from
(\ref{eqt}) in the absence of $\w_\mu$, by replacing $\phi/(6\xi^2)\ra M_p^2$.
Eq.(\ref{ff})
is consistent with the equation $\phi^2\!=\!-\tilde R$  introduced
to linearise (\ref{inA}) into (\ref{alt}).
To see this, apply (\ref{WGS}) to $\phi^2\!=\!-\tilde R$, which
becomes $\langle\phi^2\rangle\!=\!- \tilde R$;
after decoupling of massive $\w_\mu$ this gives $\langle\phi^2\rangle\!=\!- R$.
With notation  $\Lambda\!=\!\langle\phi^2\rangle/4$),
then $4\Lambda=-R$, in agreement with (\ref{ff}).

\bigskip
\def\theequation{C-\arabic{equation}}
\def\thesubsection{C}
\setcounter{equation}{0}
\def\thefigure{C-\arabic{figure}}

\subsection{Higgs sector: $\cL_H$ and the matrix $\cM^2(\sigma)$ }
\label{AB}

For convenience, we write here in the Riemannian notation and in the symmetric phase
the form of $\cL_H$ shown in the text in the  Weyl geometry notation
eq.(\ref{higgsR}) after step (\ref{rep})
\medskip
\bea\label{action2}
\cL_H\!\!\!\!&=&\!\!\!\sqrt{g}\,\Big\{\frac{-1}{2}\Big[ \frac16 \theta^2 R +(\partial_\mu\theta)^2
-\frac{\q}{2}\, \nabla_\mu (\theta^2 \w^\mu)\Big]
-\frac{1}{\eta^2} C_{\mu\nu\rho\sigma}^2
+\frac18 \,\q^2\,\theta^2\, \Big[\w_\mu
-\frac{1}{\q}\,\nabla_\mu \ln\theta^2\Big]^2
-V
\nonumber\\[6pt]
&+&
\!\!\!
\frac12 \big\vert (\partial_\mu- \,\q/2\,\w_\mu)\, h\,\vert^2 +\frac12\, h^2\cA_\mu \cA^\mu
-\frac14
\Big[ \frac{1}{\gamma^2}\,F_{\mu\nu}^2 +2\sin\chi\,F_{\mu\nu} F_y^{\mu\nu} +F_{y\,\mu\nu}^2\Big]\Big\},
\eea

\medskip\noindent
where 
$\theta^2=(1/\xi^2)\,\phi^2+\xi_h\, h^2$ denotes the radial direction in the fields space
with
\bea\label{vv}
V=\frac{1}{4!}\Big[ 6 \lambda\, h^4+ \xi^2 (\theta^2 -\xi_h h^2)^2\Big],
\eea
and $\langle\theta\rangle^2=6 M_p^2$.
The first line in $\cL_H$ is similar to
that of a single field case, see eq.(\ref{ST}) for  $\theta^2 \leftrightarrow (1/\xi^2)\,\phi^2$.
Note that $\cL_H$ is invariant under the Weyl gauge transformation eq.(\ref{WGS})
(one checks that the first square bracket is invariant,
while for the remaining terms this is easily verified).
From this action eq.(\ref{iL})  then follows, via a Stueckelberg mechanism.

In the formal limit when the radial direction in field space
$\langle\theta\rangle\ra 0$ ($M_p\ra 0$) which restores the Weyl gauge symmetry,
then from the definition of $\theta$ we see
that $\phi\ra 0$ and $h\ra 0$ (EW symmetry is also restored) and
therefore the potential vanishes
$V\ra 0$, as expected due to the Weyl gauge symmetry.

\bigskip
The Higgs-dependent matrix $\cM^2(\sigma)$ introduced in  eq.(\ref{massmatrix})
in  basis $X\!=\!(B'_\mu, A_\mu^3, \w'_\mu)$ is
\smallskip
\bea\label{mass1}
\cM^2\!(\sigma)\!=\!
\frac{3 M_p^2}{2} \sinh^2\!\frac{\sigma}{M_p \sqrt 6}\!
\begin{pmatrix}
  g^{' 2} & - g g'  & - g^{' 2} \tan\tilde\chi \\
  - g g'  & g^2     &  g g' \tan\tilde\chi\\
  - g^{' 2} \tan\tilde\chi & g g' \tan\tilde\chi & g^{' 2} \tan^2\tilde\chi
  \!+\! \q^2 \gamma^2\sec^2\tilde\chi \coth^2 \frac{\sigma}{M_p\sqrt 6}
\end{pmatrix}
\eea

\medskip\noindent
Notice that if there is
no gauge kinetic mixing $\chi\sim\tilde\chi=0$, then $\cM^2$ simplifies considerably.
The mass matrix  is diagonalised by two successive rotations of the fields;
first:
\medskip
\bea\label{mix1}
\begin{pmatrix}
  A_\mu\\
  Z_{1\,\mu}\\
  Z_{2\,\mu}
\end{pmatrix}
=\begin{pmatrix}
  \cos\theta_w & \sin\theta_w & 0\\
   -\sin\theta_w & \cos\theta_w & 0\\
   0 & 0 & 1\\
 \end{pmatrix}
\begin{pmatrix}
  B'_\mu\\
  A^3_\mu\\
  \w'_\mu
\end{pmatrix}
\eea

\medskip\noindent
After this,  $Z_1-Z_2$  mass mixing usually exists, diagonalized  by a final rotation
of suitable $\zeta$ 
\medskip
\bea\label{mix2}
\begin{pmatrix}
  A_\mu\\
  Z_\mu\\
  Z_\mu^\w
\end{pmatrix}
=\begin{pmatrix}
  1 &0 & 0\\
  0  & \cos\zeta & -\sin\zeta\\
  0  & \sin\zeta & \cos\zeta
 \end{pmatrix}
\begin{pmatrix}
  A_\mu\\
  Z_{1\,\mu}\\
  Z_{2\,\mu}
\end{pmatrix}
\eea

\medskip\noindent
Combining these two rotations we find a matrix relating
the mass eigenstates $(A_\mu, Z_\mu, Z_\mu^\w)$ to the
gauge eigenstates $X_\mu=(B'_\mu, A_\mu^3, \w'_\mu)$. The inverse of this matrix
is shown in eq.(\ref{U}).

\vspace{1cm}

\bigskip
\noindent
{\bf Acknowledgement:}\,

\medskip\noindent
The author thanks Graham Ross (University of Oxford) for interesting discussions
on this work. This work was supported by a grant of the Romanian Ministry
of Education and Research, CNCS-UEFISCDI,
project number PN-III-P4-ID-PCE-2020-2255. % (PNCDI~III).

%\newpage

\end{document}